\def\commentType{0}

\def\anonpaper{0}

\if\anonpaper1
\documentclass[acmtog,anonymous=true,review=true,screen]{acmart}
\renewcommand\footnotetextcopyrightpermission[1]{} 
\else
\documentclass[acmtog,anonymous=false,review=false,screen]{acmart}
\fi

\usepackage[skip=5pt]{caption}

\def\BibTeX{{\rm B\kern-.05em{\sc i\kern-.025em b}\kern-.08emT\kern-.1667em\lower.7ex\hbox{E}\kern-.125emX}}

\acmJournal{TOG}
\acmYear{2022}\acmVolume{XX}\acmNumber{X}\acmArticle{XX}\acmMonth{1} \acmDOI{}

\acmSubmissionID{485}

\usepackage[utf8]{inputenc}
\DeclareUnicodeCharacter{2212}{-}

\usepackage{csquotes}
\usepackage{amsmath}
\usepackage{empheq}

\usepackage[]{todonotes}                     

\usepackage{xargs}                           
\usepackage{url}                             
\usepackage[most]{tcolorbox}                 
\usepackage[nameinlink,capitalise]{cleveref} 
\usepackage{microtype}                       
\usepackage{wrapfig}                         
\usepackage{booktabs}                        
\usepackage{tabularx}
\usepackage{cancel}                          
\usepackage{cases}                           
\usepackage{mathtools}                       
\usepackage[outline]{contour}                
\usepackage{xfrac}                           
\usepackage{cutwin}                          
\usepackage{shapepar}
\usepackage{units}                           
\usepackage{pgf}
\usepackage{braket}                          
\usepackage{accents}
\MakeRobust{\underaccent} 
\usepackage{import}

\usepackage{enumitem}                        
\usepackage{xr}
\makeatletter
\newcommand*{\addFileDependency}[1]{
  \typeout{(#1)}
  \@addtofilelist{#1}
  \IfFileExists{#1}{}{\typeout{No file #1.}}
}
\makeatother

\newcommand*{\myexternaldocument}[1]{
    \externaldocument{#1}
    \addFileDependency{#1.tex}
    \addFileDependency{#1.aux}
}

\usepackage{pgfplots}
\pgfplotsset{compat=newest}
\usepgfplotslibrary{groupplots}
\usepgfplotslibrary{dateplot}

\newcolumntype{Y}{>{\centering\arraybackslash}X}

\usepackage[ruled]{algorithm2e} 

\SetAlFnt{\small}
\SetAlCapFnt{\small}
\SetAlCapNameFnt{\small}
\SetAlCapHSkip{0pt}

\crefname{algocf}{alg.}{algs.}
\Crefname{algocf}{Algorithm}{Algorithms}
\crefname{section}{Sec.}{Secs.}
\Crefname{section}{Section}{Sections}

\newtcbox{\mymath}[1][]{%
    nobeforeafter, math upper, tcbox raise base,
    enhanced, colframe=blue!30!black,
    colback=blue!30, boxrule=1pt,
    #1}

\tcbset{
    colback=white!99!black,
    colframe=white!75!black,
    boxrule=0.15mm,
    arc=0.3mm,
    boxsep=0pt,left=2pt,right=4pt,top=2pt,bottom=4pt
}

\newcommandx{\note}[2][1=]{\todo[inline,linecolor=lightgray,backgroundcolor=lightgray!25,bordercolor=lightgray,#1]{\textbf{Note} #2}}
\newcommandx{\precondition}[2][1=]{\todo[inline,linecolor=violet,backgroundcolor=violet!25,bordercolor=violet,#1]{\textbf{Required Knowledge} #2}}
\newcommandx{\postcondition}[2][1=]{\todo[inline,linecolor=teal,backgroundcolor=teal!25,bordercolor=teal,#1]{\textbf{Takeaway} #2}}

\ifnum\commentType=0
    \newcommandx{\customComment}[3]{}
    \newcommandx{\customTODO}[3]{}
\fi
\ifnum\commentType=1
    \newcommandx{\customComment}[3]{{\footnotesize\textcolor{#2}{[\textsl{#1: #3}]}}}
    \newcommandx{\customTODO}[3]{{\footnotesize\textcolor{#2}{[\textsl{#1: #3}]}}}
\fi
\ifnum\commentType=2
    \usepackage{pdfcomment}
    \newcommandx{\customComment}[3]{\pdfcomment[icon=Comment,opacity=0.5,color=#2,author=#1]{#3}}
    \newcommandx{\customTODO}[3]{\pdfcomment[icon=Note,opacity=0.5,color=#2,author=#1]{#3}}
\fi
\ifnum\commentType=3
    \usepackage{pdfcomment} 
    \usepackage{silence}
    \newcommandx{\customComment}[3]{\todo[color=#2!40,size=\small]{\textbf{#1:} #3}}
    \newcommandx{\customTODO}[3]{\todo[color=#2!40,size=\small]{\textbf{#1:} #3}}
\fi

\definecolor{amber}{rgb}{1.0, 0.49, 0.0}
\definecolor{darkgreen}{rgb}{0.1, 0.7, 0.1}
\newcommandx{\All}[1]{\customComment{All}{red}{#1}}
\newcommandx{\wojtek}[1]{\customComment{Wojtek}{blue}{#1}}
\newcommandx{\keenan}[1]{\customComment{Keenan}{amber}{#1}}
\newcommandx{\rohan}[1]{\customComment{Rohan}{brown}{#1}}
\newcommandx{\dario}[1]{\customComment{Dario}{darkgreen}{#1}}

\newcommandx{\anon}[1]{\customComment{Authors}{amber}{#1}}

\newcommandx{\figurelist}[1]{\customComment{\textbf{Figures}}{red}{
\begin{itemize}
    #1
\end{itemize}}}

\newcommand{\expect}[1]{\mathbb{E}\left[#1\right]}
\DeclarePairedDelimiter{\abs}{\lvert}{\rvert}

\newcommand*\diff{\mathop{}\!\mathrm{d}}
\newcommand{\euler}{\mathrm{e}}

\newcommand*{\ball}{B}
\newcommand*{\surf}{{\partial \ball}}

\newcommand*\probSurf{{\mathbb{P}}^\surf}
\newcommand*\probBall{{\mathbb{P}}^\ball}

\newcommand*\pdfSurf{p^\surf}
\newcommand*\pdfBall{p^\ball}

\newcommand{\est}[1]{\widehat{#1}}

\newcommand*{\bc}{g}
\newcommand*{\src}{f}

\newcommand*{\pathtr}{T}
\newcommand*{\pathth}{W}
\newcommand*{\pathsrc}{S}

\newcommand*{\so}{\alpha}
\newcommand*{\fo}{\vec{\omega}}
\newcommand*{\zo}{\sigma}

\newcommand*{\zoh}{\zo'}
\newcommand*{\srch}{\src'}
\newcommand*{\bch}{\bc'}

\newcommand*{\uCtrl}{\bar{\zo}}

\newcommand{\wwLo}{w^-}
\newcommand{\wwHi}{w^+}

\newcommand{\secref}[1]{Section~\ref{sec:#1}}

\newcommand{\eg}{e.g.}
\newcommand{\ie}{i.e.}

\newcommand{\ala}{\emph{\`{a} la}}

\myexternaldocument{supplemental}

\begin{document}



\title{Grid-Free Monte Carlo for PDEs with Spatially Varying Coefficients}

\author{Rohan Sawhney$^*$}
\email{rohansawhney@cs.cmu.edu}
\affiliation{%
  \institution{Carnegie Mellon University}
  \country{USA}
}

\author{Dario Seyb$^*$}
\email{dario.r.seyb.gr@dartmouth.edu}
\affiliation{%
  \institution{Dartmouth College}
  \country{USA}
}

\author{Wojciech Jarosz$^{\dagger}$}
\email{wjarosz@dartmouth.edu}
\affiliation{%
  \institution{Dartmouth College}
  \country{USA}
}

\author{Keenan Crane$^{\dagger}$}
\email{kmcrane@cs.cmu.edu}
\affiliation{%
  \institution{Carnegie Mellon University}
  \country{USA}
}

\newcommand\extrafootertext[1]{%
    \bgroup
    \renewcommand\thefootnote{\fnsymbol{footnote}}%
    \renewcommand\thempfootnote{\fnsymbol{mpfootnote}}%
    \footnotetext[0]{#1}%
    \egroup
}


\begin{abstract}
   Partial differential equations (PDEs) with spatially-varying coefficients arise throughout science and engineering, modeling rich heterogeneous material behavior.  Yet conventional PDE solvers struggle with the immense complexity found in nature, since they must first discretize the problem---leading to spatial aliasing, and global meshing/sampling that is costly and error-prone.  We describe a method that approximates neither the domain geometry, the problem data, nor the solution space, providing the exact solution (in expectation) even for problems with extremely detailed geometry and intricate coefficients.  Our main contribution is to extend the \emph{walk on spheres (WoS)} algorithm from constant- to variable-coefficient problems, by drawing on techniques from volumetric rendering.  In particular, an approach inspired by \emph{null-scattering} yields unbiased Monte Carlo estimators for a large class of 2nd-order elliptic PDEs, which share many attractive features with Monte Carlo rendering: no meshing, trivial parallelism, and the ability to evaluate the solution at any point without solving a global system of equations.



\end{abstract}


\begin{teaserfigure}\label{fig:teaser}
   \centering
   \includegraphics{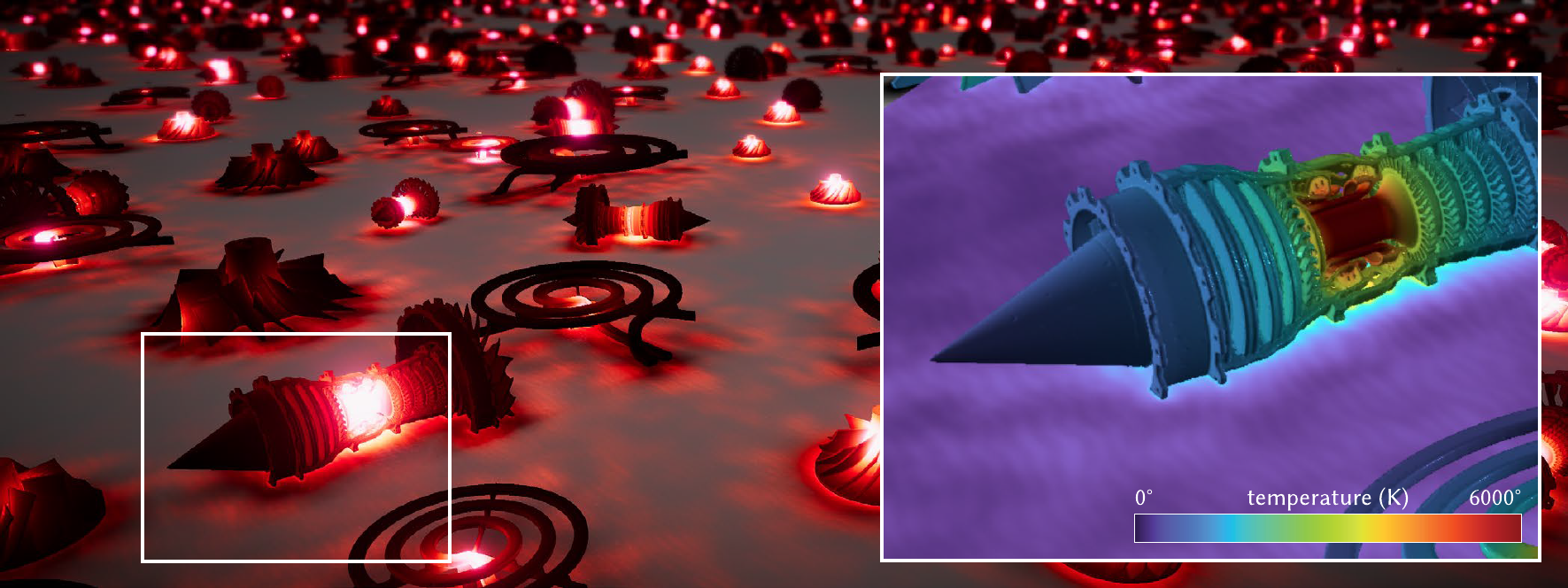}
   \caption{Distribution of heat (inset) radiating from infinitely many blackbodies---about 600M effective boundary vertices are visible from this viewpoint alone. (Here we visualize a 2D slice of the full 3D solution.)  Our Monte Carlo PDE solver directly captures fine geometric detail and intricate spatially-varying coefficients without meshing, sampling, or homogenizing the 3D domain, by building on techniques from volumetric rendering.\label{fig:teaser}}
\end{teaserfigure}

\maketitle

\extrafootertext{The symbols $^*$ and $^\dagger$ indicate equal contribution.}

\section{Introduction}\label{sec:intro}


PDEs with spatially-varying coefficients describe a rich variety of phenomena. In thermodynamics, for example, variable coefficients model how heterogeneous composite materials conduct or insulate heat.  Much as early algorithms for photorealistic rendering were motivated by predictive lighting design~\cite{ward1998rendering}, such models can be used to predict and improve thermal efficiency in building design~\cite{zalewski2010experimental}.  Likewise, variable permittivity in electrostatics impacts the design of antennas~\cite{ozdemir2005variable} and the simulation of biomolecules~\cite{fahrenberger2014simulation}; in hydrology, variable transmissivity of water through soil impacts remediation strategies for groundwater pollution~\cite{willmann2010coupling}.  More directly connected to our work, variable coefficients in the light transport equation are used to model heterogeneity in participating media~\cite{Novak:2018:Monte}.  Beyond spatially-varying materials, variable coefficients can also be used to model curved geometry, by using PDE coefficients on a flat domain to encode an alternative \emph{Riemannian metric} (see \cref{sec:WalksOnCurvedSurfaces}).

\begin{figure}[t!]
    \centering
    \footnotesize
    \includegraphics{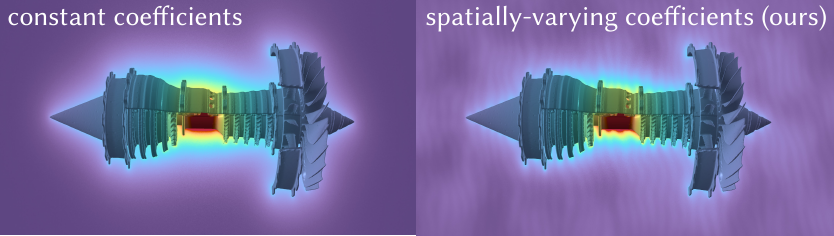}%
    \caption{We directly resolve the detailed effects of, e.g., spatially-varying material density---without resorting to homogenization of PDE coefficients.}
    \label{fig:varying-vs-constant}
\end{figure}

\begin{figure}[b!]
   \includegraphics{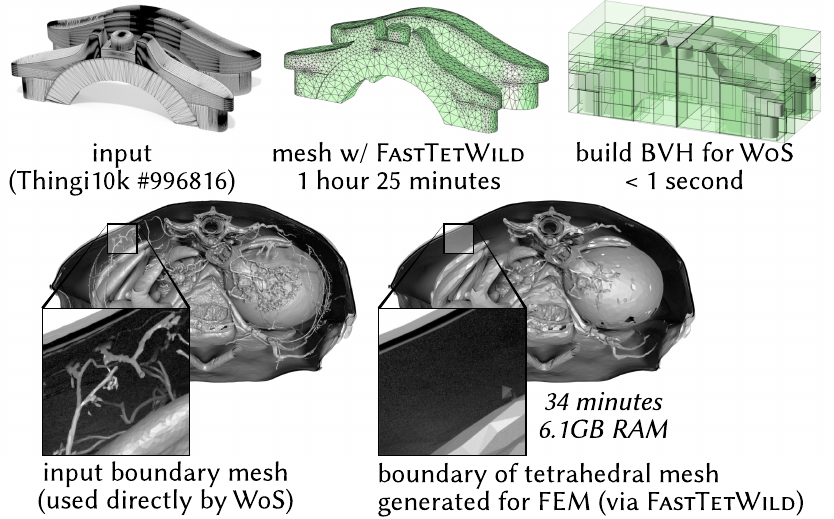}
   \caption{\emph{Top:} The bottleneck in conventional methods is often not the solve itself, but rather the cost of meshing (here, via \citet{Hu:2020:fTetWild}). As in rendering, WoS needs only build a simple bounding volume hierarchy (BVH). \emph{Bottom:} Conventional methods also sacrifice spatial detail---here destroying key features like blood vessels. \emph{Figures from \citet{Sawhney:2020:Monte}.}\label{fig:discretization-challenges}}
\end{figure}

Our method computes the exact solution (in expectation) to 2nd-order linear elliptic equations of the form%
\begin{equation}
  \label{eq:pde}
  \arraycolsep=3pt
  \boxed{
      \begin{array}{rcrl}
         \nabla \cdot (\so({x}) \nabla u) + \fo({x}) \cdot \nabla u - \zo({x}) u &=& -\src({x}) &\text{ on } \Omega , \\
         u &=& \bc({x}) & \text{ on } \partial \Omega.
      \end{array}
  }
\end{equation}
%
%
Here \(\Omega\) is a region in \(\mathbb{R}^n\), $\so: \Omega \mapsto \mathbb{R}_{> 0}$ and $\fo: \Omega \mapsto \mathbb{R}^n$ are a twice-differentiable function and vector field (resp.), and $\zo: \Omega \mapsto \mathbb{R}_{\geq 0}$ is continuous (\(C^0\)). These conditions are sufficient to ensure ellipticity. The \emph{source term} $\src: \Omega \mapsto \mathbb{R}$ and \emph{boundary values} $\bc: \partial \Omega \mapsto \mathbb{R}$ need not be continuous. \cref{fig:pde-coeffs} illustrates the effect of each term on the solution \(u\); see \cref{sec:differential-equations} for further background on PDEs.  Note that unlike methods for \emph{numerical homogenization}~\cite{Desbrun:2013:MAS}, we aim to directly resolve the original, detailed solution (\cref{fig:varying-vs-constant}).

Though many methods are available for solving such PDEs, they all suffer from a common problem: the need to spatially discretize (e.g., mesh or point-sample) the domain interior.  Even so-called \emph{meshless methods (MFEM)} must carefully distribute interior nodes; \emph{boundary element methods (BEM)} must be integrated with volumetric methods to handle interior terms (see \cref{sec:RelatedWorkAndComparisons}).  For problems with intricate geometry (as in \cref{fig:teaser}), discretization is hence a major burden on designers, scientists, and engineers: even state-of-the-art methods are error-prone, can take hours of preprocessing, and can destroy application-critical features due to spatial aliasing (\cref{fig:discretization-challenges}).  Such problems are further compounded by variable coefficients, since now the discretization must also be carefully adapted to regions where coefficients exhibit fine detail (\cref{fig:amr}).  Moreover, many meshing and sampling algorithms for constant-coefficient problems do not work out of the box on spatially-varying problems.

\begin{figure}[t!]
   \vspace{-\baselineskip}\includegraphics{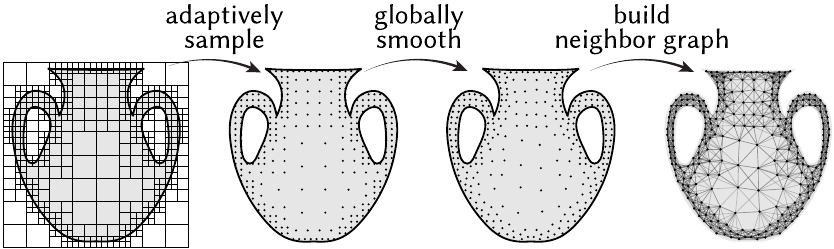}
   \caption{So-called ``meshless'' methods still perform a process akin to global meshing, which can result in spatial aliasing of fine features.  One ends up with a mesh-like structure which must satisfy stringent sampling criteria to avoid numerical blowup, and must still solve a large globally-coupled linear system. (Figure adapted from \protect{\cite[Figure 6]{Pauly:2005:MAF}}.)\label{fig:MeshlessMeshing}}
\end{figure}

Overall, these challenges make it difficult (if not impossible) to analyze large, heterogeneous systems (as in \cref{fig:teaser}) which commonly arise in real applications---say, directly analyzing a \emph{building information model (BIM)} that includes heating ducts, plumbing, insulation, etc., rather than a simplified geometric proxy.  Yet scenes of this size and complexity are commonplace in rendering.  Why such a big gap between our ability to \emph{visualize} and \emph{analyze} complex scenes?  A major reason is that rendering has moved away from methods like \emph{finite element radiosity}~\cite{Cohen:1993:Radiosity}, and toward Monte Carlo methods, both to handle more intricate light transport phenomena---and to avoid difficult meshing problems~\cite[Chapter 1]{WannJensen:2001:MCR}.  For PDEs however, Monte Carlo techniques have received comparatively little attention.

An attractive alternative are \emph{grid-free Monte Carlo methods}, which solve PDEs \emph{without} discretizing the problem domain (\cref{sec:stochastic-methods}).  The starting point is the \emph{walk on spheres (WoS)} method of \citet{Muller:1956:Continuous}, which applies a recursive integral formulation akin to the classic \emph{rendering equation}~\cite{Kajiya:1986:Rendering}.  This approach side-steps many challenges faced by conventional solvers: it can evaluate the solution at any point without meshing the domain or solving a global system, is trivial to parallelize, and works directly with any boundary representation (implicit surfaces, spline patches, etc.), including low-quality meshes not designed for finite-element analysis.  Following \citet{Sawhney:2020:Monte}, recent work in computer graphics explores how to generalize and accelerate WoS, by drawing inspiration from geometry processing and Monte Carlo rendering \citep{Nabizadeh:2021:InfDomain,Mossberg:2021:GAM,Marschner:2021:SOS,Krayer:2021:HPD}.  To date, however, WoS still handles only a small class of constant-coefficient PDEs, limiting its use in applications.

\subsection{Contributions}

We generalize WoS to a large set of variable-coefficient PDEs, by establishing a link with recent \emph{null-scattering} techniques for rendering heterogeneous participating media~\citep{Novak:2018:Monte}.  To our knowledge, there is no other known way to solve such PDEs without spatially discretizing the problem domain (whether by previous WoS methods, or any other means).  These methods are appropriate for problems involving the steady-state of a diffusive process (in contrast to, say, the dynamics of large scale deformations).  Specifically, we provide:

\begin{itemize}
    \item A novel reformulation of variable-coefficient 2nd-order linear elliptic PDEs amenable to Monte Carlo methods.
    \item Efficient WoS algorithms for these PDEs, inspired by the \emph{delta tracking} \citep{Woodcock:1965:Techniques} and \emph{next-flight} \cite{Cramer:1978:Application} methods from volume rendering.
    \item A variance reduction strategy that significantly reduces noise in problems with high-frequency coefficients, based on insights from neutron transport.
\end{itemize}

\noindent In the process, we obtain a precise mathematical picture of the relationship between diffusive variable-coefficient PDEs and heterogeneous participating media.  Since we transform the original problem into a constant-coefficient PDE with a modified source term, we get the usual convergence guarantee for WoS algorithms: the variance of an \(N\)-sample estimate decreases at a rate \(1/N\), with a negligible amount of bias due to the \(\varepsilon\)-shell (see \cite[Section 6.1]{Sawhney:2020:Monte}).  Numerical experiments on several thousand models empirically verify convergence behavior (see \cref{fig:solvers-convergence,fig:nf-convergence,fig:solution-convergence-ww,fig:discrete-bias}).

\paragraph{Limitations.} WoS is still an emerging class of methods that does not yet support all the features of more mature solvers (such as FEM)---for example, it is still not known how to handle general Neumann or Robin boundary conditions (as required for, e.g., linear elasticity). These open problems are largely orthogonal to the questions we address here, and are left to future work.  See \cref{sec:LimitationsAndFutureWork} for further discussion.

\section{Background}
\label{sec:background}

The derivation of our method depends on concepts from PDE theory, the theory of integral equations, stochastic calculus, and volumetric rendering.  Since we expect few readers to be familiar with all of these topics, we provide essential background here.
\cref{sec:walk-on-spheres} then describes the basic WoS algorithm, which is the starting point for our variable-coefficient algorithm in \cref{sec:our-method}.  For a gentler introduction to WoS, see \citet[Section 2]{Sawhney:2020:Monte}.

\subsection{Notation}
\label{sec:notation}

For any region $A \subset \mathbb{R}^n$, we use $|A|$ to denote its volume and $\partial A$ to denote its boundary. For any point $x \in A$, $\overline{x}$ denotes the point closest to ${x}$ on the boundary $\partial A$. Throughout, $\Omega \subset \mathbb{R}^n$ denotes the domain of interest, $\ball({x})$ is a ball centered on \(x\) and contained in $\Omega$, and $\partial \Omega_{\epsilon} \coloneqq \{{x} \in \Omega : |{x} - \overline{{x}}| < \epsilon\}$ denotes an \emph{epsilon shell} around $\partial \Omega$. We use $\vec{u}$ for a vector field on \(\mathbb{R}^n\), and $\nabla$ and $\nabla \cdot$ for the gradient and divergence operators (resp.), so that $\Delta \coloneqq \nabla \cdot \nabla$ is the \emph{negative-semidefinite} Laplacian. We use $\mathcal{U}$ to denote the uniform distribution on \([0,1] \subset \mathbb{R}\), and $\mathcal{N}({x}, v)$ for the $n$-dimensional normal distribution with mean ${x}$ and variance $v$.  For brevity, we often omit the arguments of functions (\eg, \(f\) rather than \(f(x,y)\)).  We use \(\mathbb{E}[X]\) to denote the expected value of any random variable \(X\).

\begin{figure}[b]
    \centering
    \footnotesize
    \includegraphics{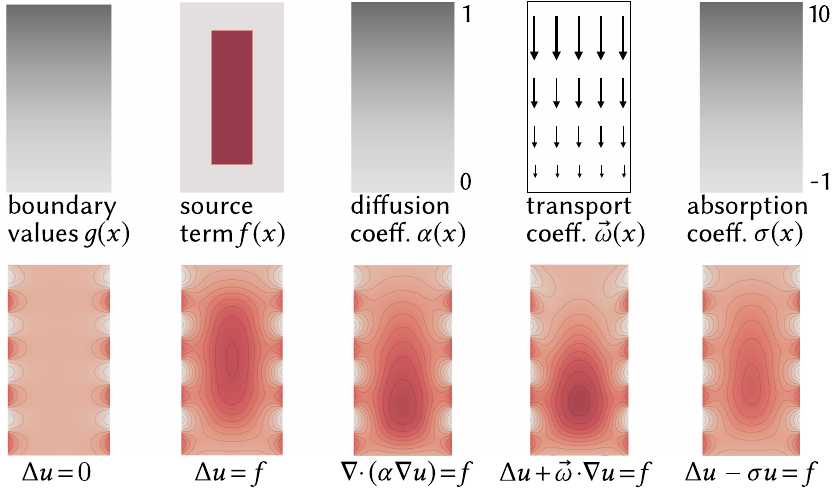}\\
    \caption{Effect of each term of \cref{eq:pde} on its solution. \emph{Left to Right}: a Laplace equation $\Delta u = 0$ smoothly diffuses boundary values $g$ into the interior. A source term $f$ adds interior contributions. The diffusion coefficient $\so({x})$ controls the diffusion rate at each point \(x\). The transport coefficient $\fo({x})$ causes values to drift along the prescribed directions. Positive/negative values of the screening coefficient \(\zo(x)\) dampen/magnify the solution, respectively.}
    \label{fig:pde-coeffs}
\end{figure}

\subsection{Differential Equations}
\label{sec:differential-equations}

A \emph{partial differential equation (PDE)} describes a function \(u\) implicitly, via relationships between derivatives in space and time.  Since this description is implicit, one must ultimately solve for an explicit function $u$ satisfying this equation---which is the \emph{raison d'\^{e}tre} for numerical PDE solvers.  A standard example is the \emph{Laplace equation}
\[
   \Delta u = \tfrac{\diff^2 u}{\diff x^2} + \tfrac{\diff^2 u}{\diff y^2} + \tfrac{\diff^2 u}{\diff z^2} = 0.
\]
Solutions to this equation, called \emph{harmonic functions}, are very smooth---describing, e.g., the steady-state distribution of heat in a room, or a smooth interpolation of color over an image (see \secref{HeterogeneousDiffusionCurves}).

\paragraph{Order and Linearity.} The \emph{order} of a PDE is the degree of its highest-order derivatives.  E.g., the Laplace equation is 2nd order in space, since it involves 2nd derivatives in \(x\), \(y\) and \(z\).  A PDE is \emph{linear} if it is a linear polynomial in the function and its derivatives. E.g., the Laplace equation is linear, whereas the \emph{inviscid Burger's equation} $\frac{\partial}{\partial t} u = -u\frac{\partial}{\partial x} u$ is nonlinear, since \(u\) is multiplied by its \(x\)-derivative.

\paragraph{Boundary conditions.} To pin down a specific solution to a PDE, we must typically prescribe \emph{boundary conditions}.  For instance, a Laplace equation with \emph{Dirichlet} boundary conditions, given by
\begin{equation}
\label{eq:laplace-equation}
\begin{array}{rcll}
   \Delta u &=& 0    & \text{ on } \Omega,\\
          u &=& g(x) & \text{ on } \partial \Omega,
\end{array}
\end{equation}
describes a smooth function \(u\) taking values \(g(x)\) at points \(x\) along the domain boundary (which describe, e.g., fixed temperature or color).  \emph{Neumann} conditions instead specify derivatives (e.g., flow of heat through the boundary).  Note that though our final algorithms handle only Dirichlet conditions, the transformations we develop in \cref{sec:our-method} make no assumptions about the type of boundary conditions---and hence could in the future be applied to Neumann problems also.

\paragraph{Source term.} Continuing with the heat analogy, a \emph{source term} \(f: \Omega \to \mathbb{R}\) adds additional ``background temperature'' to a PDE.  For instance, a \emph{Poisson equation} has the form
\begin{equation}
   \label{eq:poisson-equation}
   \begin{array}{rcll}
      \Delta u &=& -f(x) & \text{ on } \Omega,
   \end{array}
\end{equation}
possibly subject to some boundary conditions.

\paragraph{Absorption.} Finally, an \emph{absorption term} (also known as a \emph{screening term}) describes a ``cooling'' of the solution due to the background medium.  For instance, a \emph{screened Poisson equation} is given by
\begin{equation}
\label{eq:screened-poisson-equation}
   \begin{array}{rcrl}
      \Delta u - \zo u &=& -\src({x}) & \text{on}\ \Omega,\\
   \end{array}
\end{equation}
again subject to boundary conditions.

\subsection{Integral Equations}
\label{sec:integral-equations}

We can revisit the PDEs from the previous section through the lens of \emph{recursive integral equations}, akin to the classic rendering equation~\cite{Kajiya:1986:Rendering}.  The reason for doing so is that (as in rendering) solutions to these equations can be computed via recursive application of Monte Carlo integration, as we'll discuss in \cref{sec:walk-on-spheres}.

\paragraph{Boundary conditions.} In particular, the three numbered equations from the previous section are examples of so-called \emph{elliptic} PDEs---solutions to such PDEs can often be expressed as integrals.  An important example is the \emph{mean value principle}, which says that at each point $x$, the solution $u(x)$ to a Laplace equation (\cref{eq:laplace-equation}) equals the mean value of $u$ over any sphere around ${x}$:
\begin{equation}
   \label{eq:mean-value-integral}
    u(x) = \frac{1}{\abs{\partial \ball(x)}}\int_{\partial \ball(x)} u(z)\ \diff z.
\end{equation}
Notice that \cref{eq:mean-value-integral} is a recursive integral: the value of \(u\) at \(x\) depends on (unknown) values at other points \(z\).  The ``base case'' is then effectively provided by the (known) boundary values \(g\).

\paragraph{Source term.} For a PDE with a source term \(f\), such as the Poisson equation (\cref{eq:poisson-equation}), the integral representation of the solution gains a term
\begin{equation}
   \label{eq:poisson-integral}
   \int_{\ball(x)} \src(y)\ G(x,y)\ \diff y,
\end{equation}
where \(G(x,y)\) is the \emph{Green's function} for the PDE.  In general, for a linear PDE \(Lu = f\), the Green's function \(G\) satisfies \(LG = \delta\), where \(\delta\) is a Dirac delta.  When \(L\) is the Laplacian \(\Delta\), \(G\) is called the \emph{harmonic Green's function}, and describes the result of adding a single ``spike'' of heat to the source.  Importantly, the Green's function depends on the shape of the domain \(\Omega\), \eg, it will be different for a ball versus all of \(\mathbb{R}^n\).  The reason this representation is attractive is that common Green's functions are often known in closed form---we provide several explicit expressions in Sec. $1$ of the supplemental material.

\begin{figure}[t]
    \centering
    \includegraphics{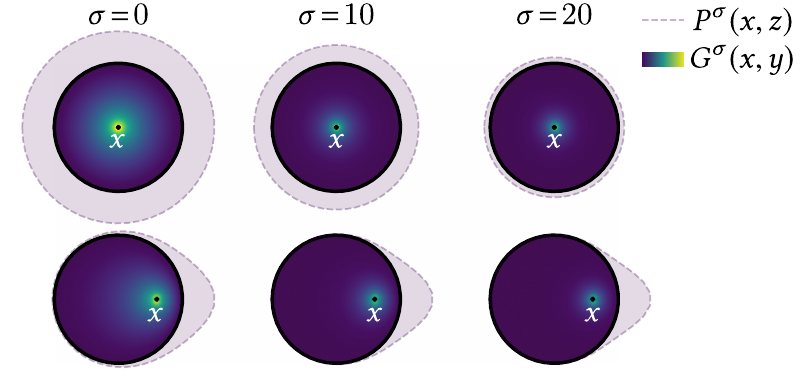}
    \caption{\emph{Top Row, Left to Right}: The Green's function $G^{\zo}({x}, {y})$ for a screened Poisson equation on a ball $\ball({x})$ becomes more localized around the point ${x}$ for increasing values of the screening coefficient $\zo$. Likewise, the magnitude of the corresponding Poisson kernel $P^{\zo}({x}, {z})$ on $\partial \ball({x})$ shrinks as $\zo$ increases. \emph{Bottom Row}: The functions $G^{\zo}$ and $P^{\zo}$ are no longer radially or angularly symmetric when ${x}$ does not lie at the center of the ball, but they are known in closed-form. In the limit as $\zo \rightarrow 0$, $G^{\zo}$ and $P^{\zo}$ become the Green's function and Poisson kernel (resp.) for the Poisson equation.}
    \label{fig:greens-functions}
\end{figure}

\paragraph{Absorption.} Adding an absorption term \(\sigma u\) to our PDE changes only the Green's function in the integral representation, which we now write as \(G^\sigma\).  \cref{fig:greens-functions} shows what this function looks like on a ball for increasing values of the coefficient $\zo$, and different points $x$ inside the ball.

\paragraph{Constant coefficients.} Putting all these terms together, the solution to a screened Poisson equation with constant coefficients (\cref{eq:screened-poisson-equation}) can be written as
\begin{equation}
   \label{eq:screened-poisson-integral}
   u(x) = \int_{\ball(c)} \src(y)\ G^{\zo}(x,y) \diff{y}\ +\ \int_{\surf(c)} u(z)\ P^{\zo}(x,z) \diff{z}.
\end{equation}
Here we also use the more general \emph{off-centered} integral formulation of the solution $u$, where the point of evaluation ${x}$ need not coincide with the center of the ball $\ball({c})$ \citep{Duffy:2015:Green,Hwang:2015:Off}.  In particular, the boundary term now incorporates the \emph{Poisson kernel} \(P^{\zo}({x}, {z})\).  Like the Green's function, the Poisson kernel describes how a ``spike'' on the boundary affects the solution, and in general is given by the normal derivative of the Green's function at the boundary (see supplemental material for explicit expressions).  For \(x=c\) and \(\zo = 0\), $P^{\zo}({x}, {z})$ reduces to $\frac{1}{\abs{\surf({x})}}$, recovering the usual mean value property.

\paragraph{Variable coefficients.} Our main variable-coefficient PDE, \cref{eq:pde}, will be elliptic as long as the diffusion coefficient $\so({x})$ is strictly positive, and the screening coefficient $\zo({x})$ is nonnegative at every point $x \in \Omega$~\cite{Evans:1998:Partial, Friedman:1975:Stochastic}. (The latter condition is in fact stronger than necessary, but it becomes difficult to check ellipticity when $\zo({x}) < 0$.)  To date, however, there appears to be no integral representation suitable for computation via WoS.  We will develop such a representation in \cref{sec:our-method}.

\subsection{Stochastic Equations}
\label{sec:stochastic-equations}

Finally, we can also express the solution to a PDE in terms of the \emph{Feynman-Kac formula} \citep[Ch. 8]{Oksendal:2003:Stochastic}, which is a fundamental result in stochastic calculus.  Considering this formulation serves two purposes: first, it is more general than (known) integral representations, providing a critical starting point for the new integral formula we develop in \cref{sec:our-method}.  Second, the Feynman-Kac formula can be put in close correspondence with the \emph{volume rendering equation (VRE)} (\cref{sec:volume-rendering}), providing us with key techniques for numerical integration.

\subsubsection{Stochastic Processes}
\label{sec:stochastic-processes}

\begin{figure}[t]
    \centering
    \begin{minipage}[c]{66pt}
    \includegraphics[width=64pt]{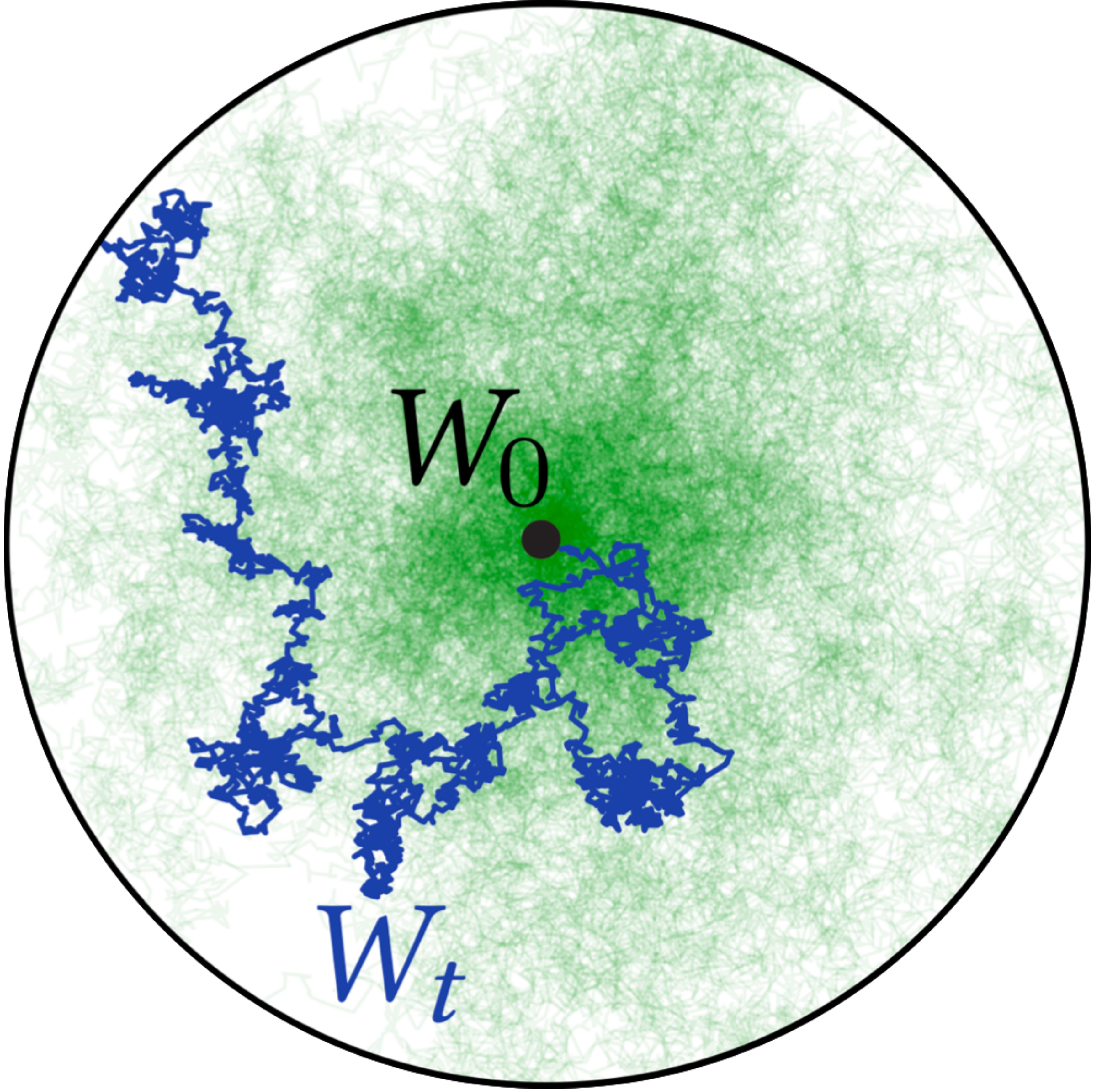}
    \end{minipage}\hfill
    \begin{minipage}[c]{174pt}
       \vspace{-\baselineskip}\caption{A \emph{Brownian motion} $W_t$ (shown in blue) describes random continuous paths in $\mathbb{R}^n$ that in aggregate (green) model diffusion. By symmetry, a Brownian random walk starting at the center of a ball exits uniformly on its surface.\label{fig:bm}}
    \end{minipage}
\end{figure}

\begin{figure}[t]
    \centering
    \includegraphics[width=\columnwidth]{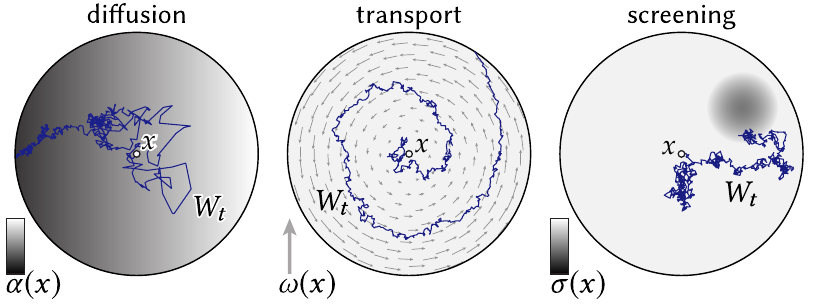}
    \caption{Walk on spheres is an acceleration technique to simulate ordinary Brownian motion $W_t$, since $W_t$ started at the center of a ball $\ball({x})$ exits uniformly on its surface. In contrast, random processes characterized by the variable coefficients $\so({x})$, $\fo({x})$ and $\zo({x})$ do not have uniform exit distributions on $\partial \ball({x})$. \emph{Left}: the diffusion coefficient $\so({x})$ scales the variance of the normal distribution from which $W_t$ is sampled, influencing the length of each increment. \emph{Center}: the transport coefficient $\fo({x})$ assigns a directionality to $W_t$’s random trajectory. \emph{Right}: the positive screening coefficient $\zo({x})$ controls the possibility of $W_t$ being absorbed inside $B({x})$; conversely $\zo({x}) < 0$ (not visualized) corresponds to the possibility of new random processes being emitted.\label{fig:VaryingCoefficients}}
\end{figure}

A \emph{continuous stochastic process} describes the trajectory of a particle taking a continuous ``random walk.''  A central example is a \emph{Brownian motion} \(W_t\), characterized by the property that increments $W_{t+s} - W_t$ follow a normal distribution \(\mathcal{N}({0}, s)\), and are independent of past values of \(W_t\) (\cref{fig:bm}).  More generally, a \emph{diffusion process} describes the trajectory of a particle moving with velocity $\fo({x})$, and subjected to random displacements of strength $\so({x})$ (\cref{fig:VaryingCoefficients}). Any such process solves a \emph{stochastic differential equation (SDE)} \citep{Higham:2001:SDESim, Kloeden:2013:NSS} of the form:
\begin{equation}\label{eq:sde}
    \diff X_t = \fo(X_t) \diff t\ +\ \sqrt{\so(X_t)}\ \diff W_t.
\end{equation}
The term $\diff W_t$ represents an infinitesimal Brownian increment that adds Gaussian ``noise'' to an otherwise deterministic trajectory.  More generally, we can imagine the particle has some chance of getting absorbed into the background medium (see \cref{fig:VaryingCoefficients}, \emph{right}); a function \(\sigma(x)\) is used to describe the strength of absorption at each point.  Absorbing random walks are quite similar in spirit to light rays in a purely absorbing homogeneous medium---as we'll discuss in \cref{sec:volume-rendering}.

\subsubsection{Feynman-Kac Formula}
\label{sec:feynman-kac-formula}

Notice that the parameters \(\alpha,\vec{\omega},\sigma\) governing the behavior of a diffusion process \(X_t\) are in 1-1 correspondence with the coefficients of our main PDE (\cref{eq:pde}).  The \emph{Feynman-Kac formula} connects these two pictures by expressing the solution to the PDE as an expectation over random trajectories of \(X_t\).

\paragraph{Boundary term.} We can start to build up the relationship between PDEs and stochastic processes by considering just a simple Laplace equation \(\Delta u = 0\) with Dirichlet boundary values \(g\) (\cref{eq:laplace-equation}).  In this case, \emph{Kakutani's principle} \citep{Kakutani:1944:TDB} states that
\begin{equation}
   \label{eq:kakutanis-principle}
   u(x) = \mathbb{E}[g(W_{\tau})],
\end{equation}
where \(\tau\) is the (random) time where \(W_t\) first hits the domain boundary \(\partial\Omega\).  In other words, the solution to a Laplace equation is just the average value ``seen'' by random walkers starting at \(x\).

\paragraph{Source term.} For a PDE with a source term, such as the Poisson equation \(\Delta u = -f\), the random walker will also pick up a source contribution of the form (see \citet[Ch. 9]{Oksendal:2003:Stochastic})
\begin{equation}
   \expect{\textstyle\int_0^\tau f(W_t)\ \diff t}.
\end{equation}
Intuitively, this value captures the average heat ``felt'' by random walkers starting at \(x\), along their overall trajectory.

\paragraph{Absorption.} To model the effect of absorption, as in a screened Poisson equation \(\Delta u - \sigma u = -f\), we can incorporate the coefficient \(\sigma\) into the boundary and source terms to get
\begin{equation}
   \label{eq:feynman-kac-screened-poisson}
   \expect{\euler^{-\sigma\tau} \bc(W_{\tau})} \qquad \text{and} \qquad \expect{\textstyle\int_0^\tau \euler^{-\sigma t} \src(W_t) \diff t},
\end{equation}
respectively \citep[Ch. 8]{Oksendal:2003:Stochastic}.  Notice that \emph{larger} values of \(\sigma\) yield \emph{smaller} contributions.  For a spatially-varying absorption coefficient, we can simply replace \(\euler^{-\sigma \tau}\) with \(\euler^{-\int_0^{\tau} \sigma(W_t) \diff t}\).

\paragraph{Feynman-Kac.} Finally, to account for variable diffusion \(\alpha(x)\) and drift \(\vec{\omega}(x)\), we can replace the Brownian motion \(W_t\) with a general diffusion process \(X_t\), \ala\ \cref{eq:sde}.  Combining the above expressions for boundary, source, and absorption terms we arrive at the full Feynman-Kac formula
\begin{equation}\label{eq:feynman-kac}
    \!\!u({x}) = \expect{\int_0^{\tau} \euler^{-\int_0^{t}\zo(X_{s}) \diff s}\ \src(X_t) \diff t\ +\ \euler^{-\int_0^{\tau}\zo(X_{t}) \diff t}\bc(X_{\tau})}\!.\!
\end{equation}

\subsubsection{Relationship Between Stochastic and Integral Equations}

The stochastic viewpoint also provides a useful interpretation of the Green's function \(G^{\zo}\) and Poisson kernel \(P^{\zo}\).  Suppose in both cases we restrict the domain to a ball \(\ball(x)\).  Then \(G^{\zo}(x,y)\) describes the (unnormalized) probability density that a random walker starting at $x$ passes through any point $y$ inside the ball; \(P^{\zo}(x,z)\) describes the probability that the walker exits through any given point $z$ on the boundary \(\partial \ball(x)\) (assuming it isn't absorbed first inside \(\ball(x)\)).  For instance, when there is no absorption (\(\zo=0\)), a Brownian motion \(W_t\) starting at the ball center \(x=c\) will exit through points on the boundary sphere $\partial \ball({x})$ with uniform probability (by symmetry).  In this case, the Poisson kernel also reduces to a constant function, in line with the mean value property (\cref{eq:mean-value-integral}).

\subsection{Volume Rendering}
\label{sec:volume-rendering}

\begin{figure}[t]
    \centering
    \includegraphics{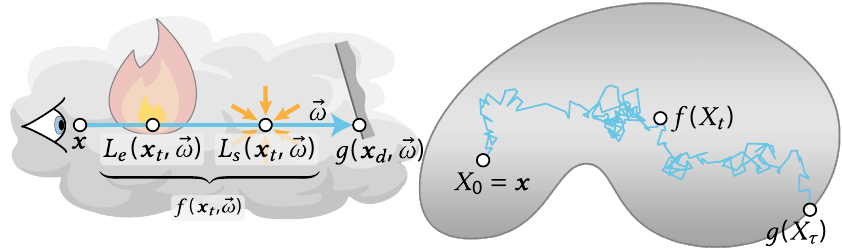}
    \caption{\emph{Left}: The radiance $L({x}, \fo)$ along a ray in the VRE is a function of the variable absorption, scattering and emission inside the medium, as well as the radiance leaving the boundary. \emph{Right}: The Feynman-Kac formula similarly determines how the source and boundary functions $\src({x})$ and $\bc({x})$ contribute to the solution of a diffusive PDE under the influence of variable coefficients, but along the trajectory of a random process $X_t$ instead.}
    \label{fig:vre-fk}
\end{figure}

In computer graphics, the radiative transport equation (RTE) \citep{Chandrasekhar:1960:Radiative} is used to describe the behavior of light in heterogeneous media that absorb, scatter and emit radiation (\cref{fig:vre-fk}, \emph{left}).  Unlike \cref{eq:pde}, the RTE is only 1st-order in space. It states that the \emph{radiance} $L({x}, \fo)$ in a medium at a point ${x}$ and along a fixed direction $\fo$ is given by:
\begin{equation}
   \label{eq:rte}
   \arraycolsep=3pt
   \begin{array}{rcrl}
      \fo \cdot \nabla L - \zo({x}) L &=& -\src({x}, \fo, L) &\text{on}\ \Omega,\\
                                    L &=&   \bc({x}, \fo, L) &\text{on}\ \partial \Omega.
   \end{array}
\end{equation}
This equation provides a recursive definition of \(L\), since the source term $\src({x}, \fo, L)$ depends on the radiance $L_s({x}, \fo)$ \emph{in-scattered} at ${x}$ (as well as any \emph{emission} $L_e({x}, \fo)$); likewise, the function $\bc({x}, \fo, L)$ describes radiance leaving the boundary. The spatially-varying \emph{extinction coefficient} $\zo({x})$ specifies the density of (scattering or absorbing) particles at ${x}$.

The integral form of the RTE is called the volume rendering equation \citep[Ch.\ 15.1]{Pharr:2016:Physically} in computer graphics. It describes the radiance $L({x}, \fo)$ from parameterized points ${x}_t \coloneqq {x} - \fo t$ along a ray of length $d$ in a heterogeneous medium:
\begin{multline}\label{eq:vre}
    L({x}, \fo) = \int_0^{d} \euler^{-\int_0^{t}\zo({x}_{s})\diff s}\ \src({x}_t, \fo, L) \diff t\ +\\ \euler^{-\int_0^{d}\zo({x}_{t}) \diff t}\bc({x}_{d}, \fo, L).
\end{multline}
Unlike the Feynman-Kac formula in \cref{eq:feynman-kac}, the VRE is deterministic. However, one faces the same challenges in estimating the VRE directly with Monte Carlo as with the Feynman-Kac formula. In particular, both na\"{i}vely resolving the \emph{transmittance function} $\euler^{-\int_0^{d}\zo({x}_{t})\diff t}$ and using explicit time-stepping to generate points ${x}_t$ along a ray lead to biased results.

\begin{figure}[t]
    \centering
    \includegraphics{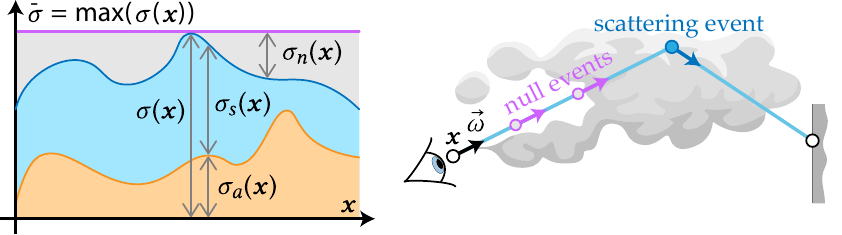}
    \caption{\emph{Left}: The delta tracking method in volume rendering artificially fills a heterogeneous medium with fictitious null matter so that the combined density is constant everywhere. \emph{Right}: Continuing forward when we probabilistically collide with null matter correctly accounts for the original heterogeneity in $\zo({x})$. The WoS algorithm in \cref{fig:wos-delta-tracking} solves variable-coefficient PDEs in a similar manner by introducing null events into the random walk.}
    \label{fig:null-events}
\end{figure}

\paragraph{Delta tracking.} The delta tracking method \citep{Woodcock:1965:Techniques, Raab:2008:Unbiased} enables unbiased Monte Carlo estimation of the VRE. Mathematically, delta tracking simply shifts the spatially-varying 0th order term $\zo({x})L$ in \cref{eq:rte} to the source term on the right-hand side \citep{Galtier:2013:Integral, Kutz:2017:Spectral}, while introducing the constant coefficient $\uCtrl$ on both sides of the equality. This yields
\begin{equation}\label{eq:rte-homogenized}
    \fo \cdot \nabla L - \uCtrl L = -\underbrace{\left(\src({x}, \fo, L) + (\uCtrl - \zo({x})) L\right)}_{=:\ \srch({x}, \fo, L)},
\end{equation}
with a corresponding integral expression:
\begin{equation}\label{eq:vre-homogenized}
    L({x}, \fo) = \int_0^{d} \euler^{-\uCtrl t}\ \srch({x}_t, \fo, L) \diff t\ +\ \euler^{-\uCtrl d}\bc({x}_{d}, \fo, L).
\end{equation}
\cref{eq:rte-homogenized,eq:vre-homogenized} are \emph{equivalent} to \cref{eq:rte,eq:vre}, but crucially, \cref{eq:vre-homogenized}  is amenable to Monte Carlo estimation since its transmittance function $\euler^{-\uCtrl t}$ does not vary spatially.
As shown in \cref{fig:null-events}, the conceptual idea is to fill a heterogeneous medium with fictitious \emph{null matter} so that the resulting medium has a constant combined density $\uCtrl \coloneqq \max(\zo({x}))$ everywhere. Doing so enables perfect, closed-form importance sampling of the transmittance by drawing samples $t$ from the probability density $\uCtrl \euler^{-\uCtrl t}$. When colliding with null matter, probabilistically continuing in the forward direction and downweighting the radiance by $\uCtrl - \zo(x)$ (see definition of $\srch$ in \cref{eq:rte-homogenized}) correctly accounts for spatial variations in $\zo({x})$.

In \cref{sec:our-method}, we derive a generalized mean value expression for the PDE in \cref{eq:pde} by applying the delta tracking transformation to the Feynman-Kac formula. This enables the design of modified WoS algorithms in \cref{sec:algorithms} inspired by techniques for path tracing heterogeneous media.

\section{Walk on Spheres}
\label{sec:walk-on-spheres}

In this section describe the basic \emph{walk on spheres algorithm}.  Walk on spheres (WoS) was originally developed by \citet{Muller:1956:Continuous} to solve the Laplace equation (\cref{eq:laplace-equation}), but has since been extended to a broader set of PDEs. \cref{sec:walk-on-spheres-algorithm} reviews WoS estimators for constant-coefficient PDEs, which serve as building blocks for the variable-coefficient extension we introduce in \cref{sec:our-method}.

\subsection{Monte Carlo Integration}\label{sec:background-mc}

WoS is a Monte Carlo estimator for the solution to a PDE.  In general, a \emph{Monte Carlo estimator} approximates an integral using random samples of the integrand.  In particular, for any (\(L^1\)) integrable function \(\phi: \Omega \to \mathbb{R}\), the quantity
\begin{align}
   I \coloneqq \int_\Omega \phi({x}) \diff{x}
\end{align}
can be approximated by the sum
\begin{align}\label{eq:mc-estimator}
   \est{I}_N \coloneqq \frac{1}{N} \sum_{i=1}^{N} \frac{\phi(X_i)}{p(X_i)}, \quad X_i \sim p,
\end{align}
where the \(X_i\) are independent random samples drawn from any probability density \(p\) that is nonzero on the support of \(\phi\).  In this paper we will express all our estimators as \emph{single-sample estimators} $\est{I}$ (dropping the subscript $N = 1$ for brevity), with the expectation that their values will be averaged over many trials to improve accuracy.

Importantly, although \(\est{I}_N\) is called an ``estimator'', it does not provide merely an estimate---rather, a well-designed estimator may give the \emph{exact} value of the integral, in expectation.  More precisely, \(\est{I}_N\) is \emph{unbiased} if \(\mathbb{E}[\est{I}_N] = I\) for \emph{any} number of samples \(N\), and \emph{consistent} if the error \(\est{I}_N - I\) goes to zero as \(N \to \infty\) with probability one~\cite[Section 1.4.4]{Veach:1997:Robust}.  Error is more often quantified by the \emph{variance} \(\text{Var}[\est{I}_N] := \mathbb{E}[(\est{I}_N - \mathbb{E}[\est{I}_N])^2]\), \ie, the average squared deviation from the expected value.  As long as \(\phi\) has finite variance, an unbiased estimator is automatically consistent (by the central limit theorem), with variance going to zero at a rate \(O(1/N)\).

\begin{figure}[t]
    \centering
    \includegraphics[width=\columnwidth]{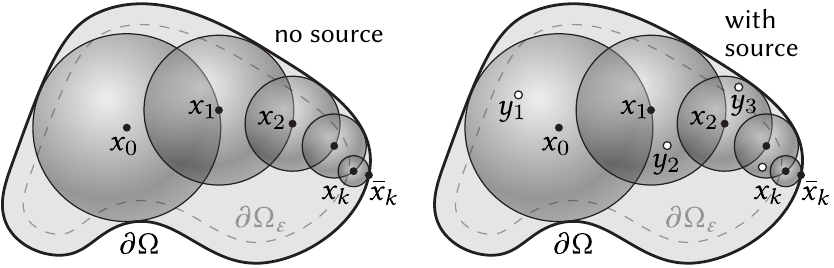}
    \caption{\emph{Left}: The walk on spheres algorithm repeatedly jumps to a random point on the largest sphere centered at the current point ${x}_k$, until it gets within an $\epsilon$ distance to the boundary. \emph{Right}: An additional random sample ${y}_{k+1}$ is required inside each ball $\ball({x}_k)$ to query the source term $f$.}
    \label{fig:wos-basic}
\end{figure}

\subsection{The Walk on Spheres Algorithm}
\label{sec:walk-on-spheres-algorithm}

Suppose we want to evaluate the solution to a basic Laplace equation \(\Delta u = 0\) with Dirichlet boundary conditions \(g\) (\cref{eq:laplace-equation}) at some point \(x_0 \in \Omega\).  The mean value formula (\cref{eq:mean-value-integral}) says that \(u(x_0)\) is equal to the average of \(u\) over any ball \(\ball(x_0) \subset \Omega\); alternatively, Kakutani's principle (\cref{eq:kakutanis-principle}) says that \(u(x_0)\) equals the expected value of \(u\) where trajectories of random walkers first hit the ball boundary:
\[
    u(x_0) = \frac{1}{\abs{\partial \ball(x_0)}}\int_{\partial \ball(x_0)} u(z)\ \diff z = \mathbb{E}[u(W_{\tau})].
\]
Both perspectives point to the same strategy for estimating \(u(x_0)\): uniformly sample a point \(x_1\) on a ball around \(x\).  If \(x_1\) is extremely close to the domain boundary (\ie, within the \(\varepsilon\)-shell \(\Omega_{\varepsilon}\)), grab the boundary value \(g(\overline{x}_1)\).  Otherwise, evaluate \(u(x_1)\).  This reasoning leads to the recursive WoS estimator
\begin{align}\label{eq:laplace-estimator}
    \est{u}(x_k) &\coloneqq \begin{cases}
    \bc(\overline{x}_k) & x_k \in \partial \Omega_{\epsilon},\\
    \est{u}(x_{k+1}) & \text{otherwise}.
    \end{cases}
\end{align}
Here the point $x_{k+1}$ is drawn from a uniform distribution on the \emph{largest} sphere centered at $x_k$, helping us reach the boundary in a small number of steps.  \cref{fig:wos-basic}, \emph{left} depicts one possible ``walk'' taken by this algorithm.  As shown in \cref{fig:discrete-bias} (and in \cite[Figure 14]{Sawhney:2020:Monte}) terminating this walk in the \(\varepsilon\)-shell introduces a negligible bias, which diminishes at a rate of \(O(1/\log \varepsilon)\) \cite{Binder:2012:RCW}.

\newcommand{\poissonWalkFigure}{%
\begin{wrapfigure}{r}{0.4\columnwidth}
    \centering
    \vspace{-1.5\baselineskip}\hspace{-2.9em}\includegraphics[scale=0.23]{figures/poisson-walk.pdf}
\end{wrapfigure}
}

\paragraph{Source term.} To incorporate a source term \(f\), we also need to estimate the integral from \cref{eq:poisson-integral} over each ball \(\ball(x_k)\) in the walk, and add this contribution to the estimate for \(u(x_k)\).  Here again we use a single-sample estimate at a point \(y_{k+1}\) (\cref{fig:wos-basic}, \emph{right}).  Though \(y_{k+1}\) could be sampled uniformly, \citet[Section 4.2]{Sawhney:2020:Monte} show that better results are obtained by importance sampling the source \(f\), Green's function \(G\), or combining strategies via \emph{multiple importance sampling}~\cite{Veach:1995:Optimally}.

\paragraph{Absorption.} Finally, to incorporate a constant absorption term \(\zo u\), as in the constant-coefficient screened Poisson equation (\cref{eq:screened-poisson-equation}), we simply need to adopt the corresponding Green's function and Poisson kernel (given in supplemental material).

Unfortunately, a generalized mean value expression like \cref{eq:screened-poisson-integral} is not readily available for PDEs with variable coefficients, making it unclear how to use WoS to solve the PDE in \cref{eq:pde}. However, the solution to this PDE can be described by the stochastic Feynman-Kac formula \eqref{eq:feynman-kac}. In the next section, we exploit the structural similarities between Feynman-Kac and the volume rendering equation \eqref{eq:vre} to derive a generalized mean value expression for \cref{eq:pde}.

\section{Our Method}
\label{sec:our-method}

\begin{figure}[t]
    \centering
    \includegraphics[width=\columnwidth]{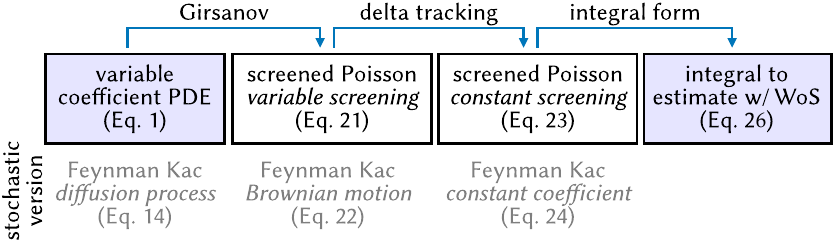}
    \caption{An overview of the transformations we apply to \cref{eq:pde} to derive an integral formulation amenable to Monte Carlo estimation with WoS.}
    \label{fig:pipeline}
\end{figure}

The key insight of our method is that while WoS cannot directly deal with spatially-varying coefficients, it can easily solve PDEs with a spatially-varying source term \(f\). Therefore, to derive a generalized mean value expression for \cref{eq:pde}, we first apply a series of transformations (shown in \cref{fig:pipeline}) that convert this PDE into an equivalent constant-coefficient screened Poisson equation. These transformations shift all spatial variability in the coefficients $\so({x})$, $\fo({x})$ and $\zo({x})$ to the source term on the right-hand side, resulting in a PDE where all the differential operators have constant coefficients (\cref{eq:pde-homogenized}). From a stochastic integral perspective, this is equivalent to reformulating the Feynman-Kac formula (\cref{eq:feynman-kac}) purely in terms of Brownian motion (instead of a generic diffusion process). We then leverage the known mean value formulation of the resulting screened Poisson PDE in \cref{eq:screened-poisson-integral} to develop Monte Carlo estimators that can be simulated with WoS.

In this section, we assume for clarity of exposition that the transport coefficient $\fo({x}) = \vec{0}$ over the entire domain. The structure of the transformations we apply remains unchanged for the $\fo({x}) \ne \vec{0}$ case, which we provide in \cref{app:stochastic-pde-derivation}. For readers not interested in the derivation, \cref{eq:pde-integral} gives an integral formulation for the solution to \cref{eq:pde}, while \cref{sec:algorithms} describes two variants of WoS to estimate this integral.

\subsection{Transformations}\label{sec:transformations}

As a first step, we apply the product rule to the 2nd-order operator in \cref{eq:pde}. We then divide the resulting equation by $\so({x})$, and apply the identity $\nabla \ln (\so({x})) = \nabla \so({x}) \left/ \so({x}) \right.$ to get:
\begin{align}\label{eq:pde-product-rule}
    \Delta u + \nabla \ln (\so({x})) \cdot \nabla u - \frac{\zo({x})}{\so({x})}u &= -\frac{\src({x})}{\so({x})}.
\end{align}
Notice that unlike the lower order terms in this intermediate equation, the 2nd-order term $\Delta u$ no longer depends on a spatially-varying coefficient.

\paragraph{Girsanov transformation.} A Girsanov transformation \citep[Ch. 8]{Oksendal:2003:Stochastic} is a powerful tool in stochastic calculus that allows a random process to be re-expressed under a change of probability measure, e.g., from a generic diffusion process $X_t$ to an ordinary Brownian motion $W_t$. As shown in \cref{app:girsanov}, applying this transformation to \cref{eq:pde-product-rule} eliminates the 1st order operator entirely, shifting all of its spatial variability into the coefficient in the lowest (0th) order term:
\begin{align}\label{eq:girsanov-pde}
    \Delta U - \zoh({x}) U &= \srch({x})\  \text{ on } \Omega,\\
    U &= \bch({x})\ \text{ on } \partial \Omega. \nonumber
\end{align}
Here,
\begin{align*}
    &U \coloneqq \sqrt{\so({x})}\ u,\quad
    \bch({x}) \coloneqq \sqrt{\so({x})}\ \bc({x}),\quad
    \srch({x}) \coloneqq \frac{\sqrt{\so({x})}}{\so({x})}\ \src({x}),\\
    &\qquad\text{and}\quad\zoh({x}) \coloneqq \frac{\zo({x})}{\so({x})} + \frac{1}{2}\left(\frac{\Delta \so({x})}{\so({x})} - \frac{|\nabla \ln (\so({x}))|^2}{2}\right).
\end{align*}
\cref{eq:girsanov-pde} is equivalent to our original PDE with variable coefficients in \cref{eq:pde}, which can be verified by substituting the expressions for $U, \bch, \srch$ and $\zoh$ back into this equation. Unlike the Feynman-Kac formula in \cref{eq:feynman-kac}, the stochastic integral expression for this PDE,
\begin{multline}\label{eq:feynman-kac-bm}
    U({x}) = \mathbb{E}\left[\int_0^{\tau} \euler^{-\int_0^{t}\zoh(W_s) \diff s} \srch(W_t) \diff t\ \right. +\\
    \left.
    \euler^{-\int_0^{\tau}\zoh(W_t) \diff t}\ \bch(W_{\tau}) \right],
\end{multline}
depends solely on the trajectories of a Brownian random walk $W_t$. We deal with the spatially-varying screening coefficient $\zoh({x})$ next.

\paragraph{Delta tracking.} At this point, the only remaining term on the left-hand side of \cref{eq:girsanov-pde} with a variable coefficient is the 0th order term $\zoh({x}) U$. We now apply the delta tracking transformation (\cref{sec:volume-rendering}) to shift this heterogeneity to the right-hand side, since we know how to express the solution of a PDE with a variable source term in integral form. In doing so, we also introduce an auxiliary coefficient $\uCtrl > 0$ into \cref{eq:girsanov-pde} by subtracting the term $\uCtrl U$ on both sides of the equality. This results in a PDE with the same structure as a screened Poisson equation:
\begin{align}\label{eq:pde-homogenized}
   \Delta U(x) - \uCtrl U(x) &= -\underbrace{\left(\srch(x) + (\uCtrl - \zoh(x)) U(x) \right)}_{=:\ \srch(x,\ U)} \text{ on } \Omega,\\
    U(x) &= \bch(x)\ \text{ on } \partial \Omega. \nonumber
\end{align}
Note that even though only constant coefficients now appear on the left-hand side, \emph{no approximation of any kind has been introduced.} However, unlike a typical PDE the solution $U$ still appears on the right-hand side (see the definition of $\srch({x}, U)$). In \cref{sec:algorithms}, we account for this dependence on $U$ in the recursive definition of our estimators by using a strictly positive value for $\uCtrl$.

Like the transformed VRE in \cref{eq:vre-homogenized} and the stochastic integrals in \cref{eq:feynman-kac-screened-poisson}, the Feynman-Kac expression for \cref{eq:pde-homogenized},
\begin{equation}\label{eq:feynman-kac-homogenized}
    U({x}) = \mathbb{E}\left[\int_0^{\tau} \euler^{-\uCtrl t} \srch(W_t, U) \diff t\ +\
    \euler^{-\uCtrl \tau}\ \bch(W_{\tau}) \right],
\end{equation}
also has a transmittance function $\euler^{-\uCtrl t}$ that no longer varies spatially.

\subsection{Generalized Mean Value Formulation}\label{sec:generalized-mean-value-formulation}

We can now express the solution $U$ of \cref{eq:pde-homogenized}, at a point ${x}$ inside a ball $\ball({c}) \subset \Omega$, using the integral formulation of the constant coefficient screened Poisson equation from \cref{eq:screened-poisson-integral}:
\begin{align}\label{eq:pde-integral-simple}
    U({x}) = \int_{\ball({c})} \!\!\!\!\srch({y}, U)\ G^{\uCtrl}({x}, {y}) \diff{y}\ +\ \int_{\surf({c})} \!\!\!\!\!\!\!\! U({z})\ P^{\uCtrl}({x}, {z}) \diff{z}.
\end{align}
As a last step, we make the substitution $U = \sqrt{\so({x})}\ u$ from \cref{eq:girsanov-pde} to express this integral in terms of the original solution variable $u$. This yields:
\begin{equation}
  \label{eq:pde-integral}
  \arraycolsep=3pt
  \boxed{
      \begin{array}{rcrl}
         u({x}) = \frac{1}{\sqrt{\so({x})}}\left( \int_{\ball({c})} \srch({y}, \sqrt{\so}\ u) \ G^{\uCtrl}({x}, {y}) \diff{y}\ \right. +\\
    \left. \int_{\surf({c})} \sqrt{\so({z})}\ u({z})\ P^{\uCtrl}({x}, {z}) \diff{z} \right).
      \end{array}
  }
\end{equation}
%
This integral can be estimated with Monte Carlo using WoS, as we show next. Unlike \cref{eq:screened-poisson-integral}, this formulation is recursively defined in both the volume and boundary integrals over the ball $\ball(c)$.

\subsection{PDE Estimation}
\label{sec:pde-estimation}

\newcommand{\branchFigure}{%
\begin{wrapfigure}{r}{0.4\columnwidth}
    \vspace{-1\baselineskip}
    \centering
    \hspace{-3em}\includegraphics[scale=1.0]{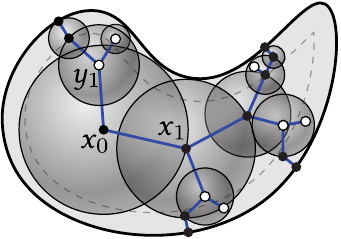}
\end{wrapfigure}
}

A single sample Monte Carlo estimate $\est{u}$ for \cref{eq:pde-integral}, at a point ${x}_k \in \ball({c})$, is given by:
\begin{multline}\label{eq:pde-estimator}
    \est{u}({x}_k) \coloneqq \frac{1}{\sqrt{\so({x}_k)}} \Bigg(\overbrace{\frac{\srch({y}_{k+1}, \sqrt{\so}\ \est{u})\ G^{\uCtrl}({x}_k, {y}_{k + 1})}{\pdfBall({y}_{k+1})\ \probBall}}^{\text{evaluate volume term with prob. } \probBall}\ +\\ \overbrace{\frac{\sqrt{\so(x_{k+1})}\ \est{u}({x}_{k+1})\ P^{\uCtrl}({x}_k, {x}_{k+1})}{\pdfSurf({x}_{k+1})\ \probSurf}}^{\text{evaluate boundary term with prob. } \probSurf}
    \Bigg),
\end{multline}
\branchFigure{}where ${y}_{k+1}$ and $x_{k+1}$ are points sampled inside and on the surface of the ball $\ball({c})$ from the probability densities $\pdfBall$ and $\pdfSurf$ (resp.). $\probBall$ and $\probSurf$ represent the probabilities with which the volume and boundary integrals are sampled. Given that the solution is recursive in both integrals, na\"{i}vely estimating both terms in \cref{eq:pde-estimator} (with, e.g., $\probBall$ = $\probSurf$ = 1) will result in an exponentially growing number of spheres to keep track of in every walk, since each sphere branches into two additional spheres (see inset). This continues until one or both of the points ${y}_{k+1}$ and $x_{k+1}$ are contained in the epsilon shell $\partial \Omega_{\epsilon}$ around the boundary. In the next section, we avoid branching walks by designing two modified WoS algorithms that vary in their choice for $\probBall$ and $\probSurf$. In Sec. $2$ of the supplemental material, we describe how to also use these algorithms to estimate the spatial gradient of \cref{eq:pde-integral}.

\section{Algorithms}\label{sec:algorithms}

The past few decades of volume rendering research have given rise to a plethora of algorithms for solving the VRE \citep{Novak:2018:Monte}. The reason for this development is that algorithms generally optimize different performance metrics to address the heterogeneity in the input coefficients, trading off between the bias, variance and speed with which the solution is estimated. As a result, the effectiveness of any one algorithm often depends on the particular coefficients in a scene. The situation is similar when solving variable-coefficient PDEs like \cref{eq:pde} with Monte Carlo. Like the unidirectional estimator in \citet[Eq. 14]{Georgiev:2019:Integral}, we provide a unified integral framework based on \cref{eq:pde-estimator} in which to develop different WoS algorithms for diffusive PDEs. In particular, we devise two variants inspired by the \emph{delta tracking} \citep{Woodcock:1965:Techniques} and \emph{next-flight} \cite{Cramer:1978:Application} methods in volume rendering. Pseudo-code for both variants is provided in Sec. $3$ of the supplemental material.

\begin{figure}[t]
    \centering
    \includegraphics[width=\columnwidth]{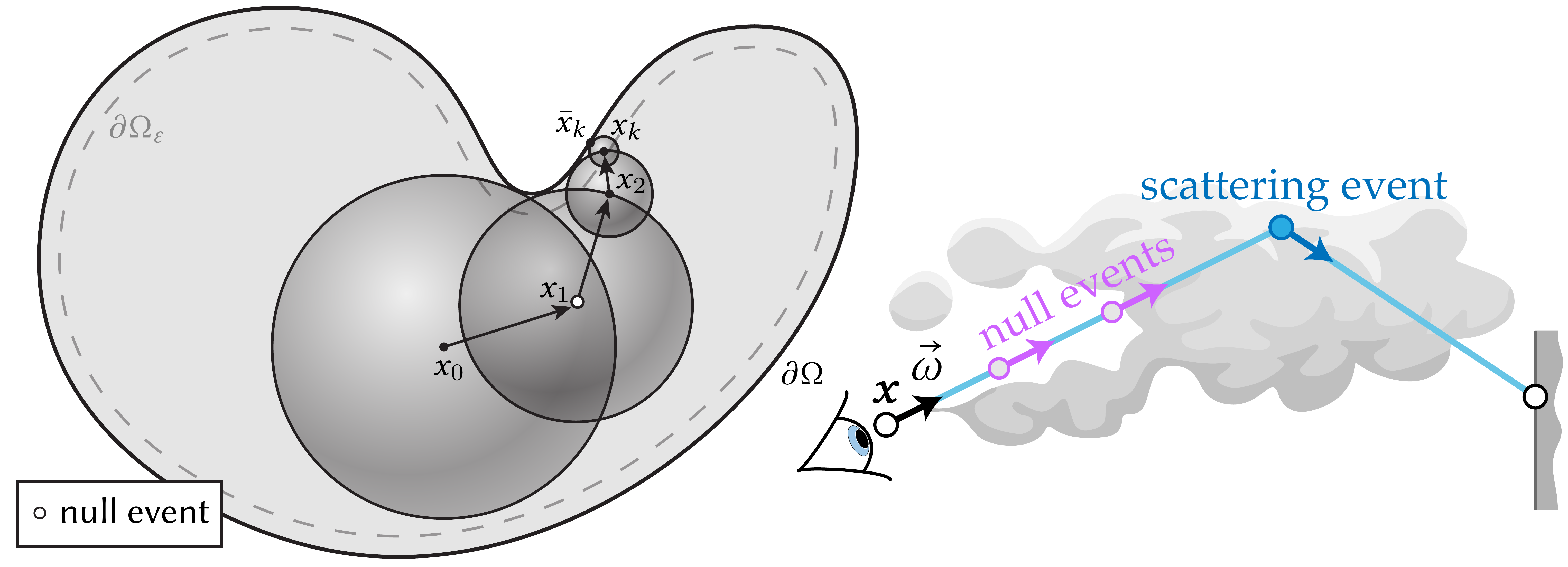}
    \caption{\emph{Left}: Unlike the standard WoS algorithm for constant coefficient problems, the delta tracking variant of WoS jumps to a random point either inside or on the surface of the largest ball $B({x}_k)$ centered at ${x}_k$. As in rendering (\emph{right}), null events sampled inside $B({x}_k)$ re-weight the solution estimate by the factor $\uCtrl - \zoh({x})$ (see definition of $\srch$ in \cref{eq:pde-homogenized}) to correctly account for the spatial variation in the problem.}
    \label{fig:wos-delta-tracking}
\end{figure}

\subsection{The Delta Tracking Variant of Walk on Spheres}\label{sec:wos-delta-tracking}

To avoid branching, the delta tracking variant of WoS takes advantage of a special property of the Poisson kernel $P^{\uCtrl}({x}_k, x_{k+1})$ \emph{for a screened Poisson equation} when ${x}_k$ lies at the center of a ball $\ball({c})$. In particular, $P^{\uCtrl}({x}_k, x_{k+1})$ equals $\frac{1 - \uCtrl \abs{G^{\uCtrl}({x}_k)}}{\abs{\surf({x}_k)}}$ for positive values of the auxiliary coefficient $\uCtrl$, with $\uCtrl \abs{G^{\uCtrl}({x}_k)}$ taking values in the range $(0, 1)$. The quantity $\abs{G^{\uCtrl}({x})}$ represents the integrated value of the Green's function over ${y}$ on $\ball({x})$ (see Sec. $1.1$ in the supplemental). One can thus use $\probSurf \coloneqq 1 - \uCtrl \abs{G^{\uCtrl}({x}_k)}$ as the probability of sampling the boundary term in \cref{eq:pde-estimator}, and $\probBall \coloneqq 1 - \probSurf$ as the probability of sampling the volume term instead. This choice results in the following non-branching procedure inside the domain $\Omega$:
\begin{equation}\label{eq:wos-delta-tracking-estimator}
    \est{u}({x}_k) \coloneqq
    \frac{1}{\sqrt{\so({{x}_k}})}
    \begin{cases}
        \frac{1}{\uCtrl}\srch({y}_{k+1}, \sqrt{\so}\ \est{u}), & \mu \sim \mathcal{U} \leq \uCtrl \abs{G^{\uCtrl}({x}_k)} \\
        \sqrt{\so(x_{k+1})}\ \est{u}(x_{k+1}) & \text{otherwise.}
    \end{cases}
\end{equation}
Here we choose the densities $\pdfBall = G^{\uCtrl}({x}_k, {y}_{k+1})\left/\abs{G^{\uCtrl}({x}_k)}\right.$ and $\pdfSurf = 1\left/\abs{\surf({x}_k)}\right.$ to importance sample ${y}_{k+1}$ and $x_{k+1}$ (resp.). If the point ${x}_k$ is contained in the epsilon shell $\partial \Omega_{\epsilon}$, then the boundary data $\bc(\overline{{x}}_k)$ is used as the value of the estimate $\est{u}({x}_k)$, and the walk is terminated; see \cref{fig:wos-delta-tracking}.

\begin{figure}[t]
    \centering
    \includegraphics{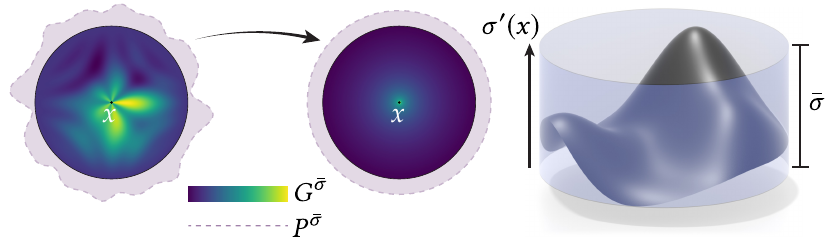}
    \caption{\emph{Left}: The Green's function and Poisson kernel on a ball $\ball({x})$ are not known in closed-form for a screened Poisson equation with a variable screening coefficient $\zoh({x})$ (\emph{right}). \emph{Center}: The delta tracking variant of WoS instead draws samples on $\ball({x})$ using the corresponding functions with a constant coefficient $\uCtrl$, and then corrects for spatial variations in the original coefficient $\zoh({x})$ by randomly sampling null events as in \cref{fig:wos-delta-tracking}.}
    \label{fig:heterogeneous-measures}
\end{figure}

The auxiliary coefficient $\uCtrl$ serves as the only free parameter in this algorithm. It can be prescribed any positive value which ensures $\probBall > 0$ (note that $\probBall = 0$ if $\uCtrl = 0$). The value it is assigned however affects the variance of the estimator. The delta tracking method for volume rendering bounds the value of the positive extinction coefficient $\zo({x})$ in \cref{eq:vre} over the entire medium by setting $\uCtrl = \max(\zo({x}))$. Numerically, this choice enables closed-form probabilistic sampling of volumetric ``events'' (absorption, scattering or null) inside the medium, as well as reflections off the boundary. We choose the parameter $\uCtrl$ in an analogous manner for the delta tracking variant of WoS by bounding the variable screening coefficient $\zoh({x})$ (which shows up inside the definition of $\srch$) from \cref{eq:pde-homogenized}. Unlike rendering, $\zoh({x})$ can take on both positive and negative values inside the domain $\Omega$ for arbitrary input coefficients $\so({x})$, $\zo({x})$ (and $\fo({x})$; see \cref{app:remove-transport}). We therefore use $\max(\zoh({x})) - \min(\zoh({x}))$ as the default value for $\uCtrl$, which enables us to use the Green's function and Poisson kernel for a constant coefficient screened Poisson equation to sample the next random point in the walk; see \cref{fig:heterogeneous-measures}.

More recent volume rendering research \citep{Novak:2014:Residual, Georgiev:2019:Integral} treats $\uCtrl$ as a control variate, rather than a bound, which is chosen based on the profile of the input coefficients. In combination with clever choices for the densities $\pdfBall$ and $\pdfSurf$, this gives rise to a broader set of more efficient algorithms compared to delta tracking. We leave such extensions to future work.

\subsection{The Next-Flight Variant of Walk on Spheres}
\label{sec:wos-next-flight}

The primary weakness of the delta tracking variant of WoS is its deteriorating performance for increasing values of the parameter $\uCtrl$; see \cref{fig:nf-walklength}. In particular, the algorithm takes shorter steps per walk as the Green's function for each ball $\ball({x})$ in \cref{eq:pde-estimator} becomes more localized with larger $\uCtrl$ (\cref{fig:greens-functions}), and the likelihood of sampling a point inside a ball becomes larger than sampling one on its surface ($\probBall > \probSurf$). As a result, the number of distance queries needed to find the largest sphere inside any domain $\Omega$ shoot up significantly. As with ray intersections for path tracing, distance queries are often the main computational bottleneck for WoS.

\begin{figure}[t]
    \centering
    \includegraphics{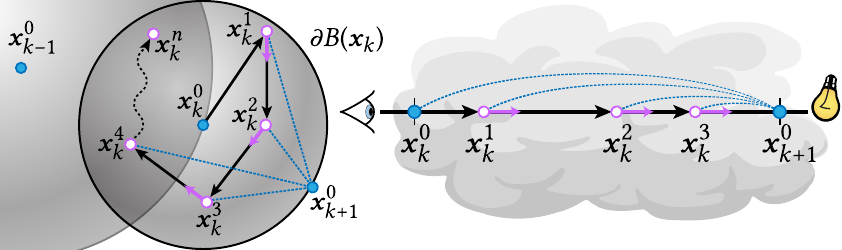}
    \caption{\emph{Left}: The next-flight variant of WoS deals with the heterogeneity in the input coefficients by evaluating \emph{off-centered} versions of the Green's function and Poisson kernel for every ball $\ball({x}^{0}_k)$ in a walk. All ``paths'' inside $\ball({x^{0}}_k)$ are required to terminate at a point ${x}^{0}_{k+1}$ on $\partial \ball({x}^{0}_k)$ to avoid branching. \emph{Right}: The next-flight method for volume rendering analogously estimates the transmittance along a ray with a predetermined end-point.}
    \label{fig:wos-path-connections}
\end{figure}

\begin{figure}[b]
    \centering
    \includegraphics[width=\columnwidth]{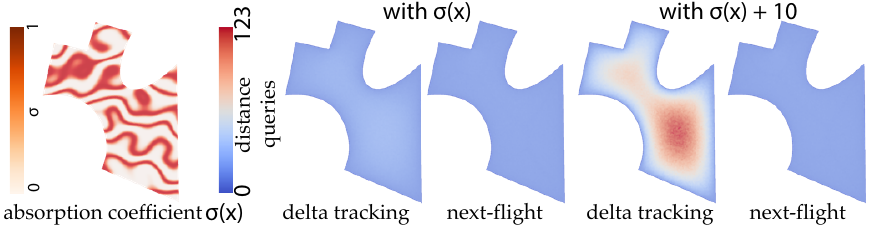}\\
    \caption{\emph{Left to Right}: The number of distance queries needed by the delta-tracking variant of WoS shoot up significantly with increasing values of the bounding parameter $\uCtrl$ (used here to bound varying absorption coefficients). In contrast, the value of $\uCtrl$ has no effect on the distance queries needed by the next-flight variant, resulting in better run-time performance.}%
    \label{fig:nf-walklength}
\end{figure}

To address this issue, we propose another variant of WoS based on the next-flight method \cite{Cramer:1978:Application} in volume rendering. At a high level, this technique always takes large steps inside $\Omega$ even when the value of $\uCtrl$ is large. As depicted in \cref{fig:wos-path-connections}, it does so by sampling a point ${x}_{k+1} \sim \pdfSurf$ on $\surf({x}_k)$ to evaluate the boundary term in \cref{eq:pde-estimator} with unit probability (i.e., $\probSurf = 1$). To compute $\est{u}({x}_k)$, it also evaluates the volume term over $\ball({x}_k)$ with probability $\probBall = 1$. However, rather than just evaluating one volume term, we expand out the recursive definition of $\est{u}$ in the volume term repeatedly over $\ball({x}_k)$ itself. This "unrolling" results in two series expressions $\est{\pathtr}$ and $\est{\pathsrc}$ containing contributions from both the boundary and volume terms (see below). Reusing the same sample point ${x}_{k+1}$ across all entries in $\est{\pathtr}$ yields a new single point estimate for $\est{u}({x}_k)$:
\begin{equation}\label{eq:next-flight-estimator}
    \est{u}({x}_k) \coloneqq \frac{1}{\sqrt{\so({x}_k)}}
    \left(\sqrt{\so({x}_{k+1})}\ \est{u}({x}_{k+1})\ \est{\pathtr}({x}_k, {x}_{k+1})\ +\ \est{\pathsrc}({x}_k)
    \right),
\end{equation}
where
\begin{align*}
    \est{\pathtr}({x}^0, {z}) &\coloneqq \sum_{j=0}^{M}\frac{P^{\uCtrl}({x}^j, {z})}{\pdfSurf({z})}\ \prod^{j-1}_{l=0} \pathth(l),\\
    \est{\pathsrc}({x}^0) &\coloneqq \sum_{j=1}^M \frac{\src({x}^j)}{\sqrt{\so({x}^j)}\ (\uCtrl - \zoh({x}^j))}\ \prod^{j-1}_{l=0} \pathth(l),\\
    \pathth(l) &\coloneqq \frac{G^{\uCtrl}({x}^l, {x}^{l+1})\ (\uCtrl - \zoh({x}^{l+1}))}{\pdfBall({x}^{l+1})}.
\end{align*}
As before, the subscript $k$ differentiates the steps in a walk, while the superscript $l$ indicates the points ${x}^l \sim \pdfBall$ sampled inside $\ball({x}_k)$ to evaluate the series $\est{\pathtr}({x}^0, {z})$ and $\est{\pathsrc}({x}^0)$. The number of terms $M$ in both series is not predetermined\textemdash rather the running ``throughput'' value $\prod_{l=0}^{j-1} \pathth(l)$ is used as a \emph{Russian Roulette} probability to determine $M$. Branching is avoided by reusing the same boundary estimate $\est{u}({x}_{k+1})$ across all entries in $\est{\pathtr}$.

\begin{figure}[b]
    \centering
    \includegraphics[width=\columnwidth]{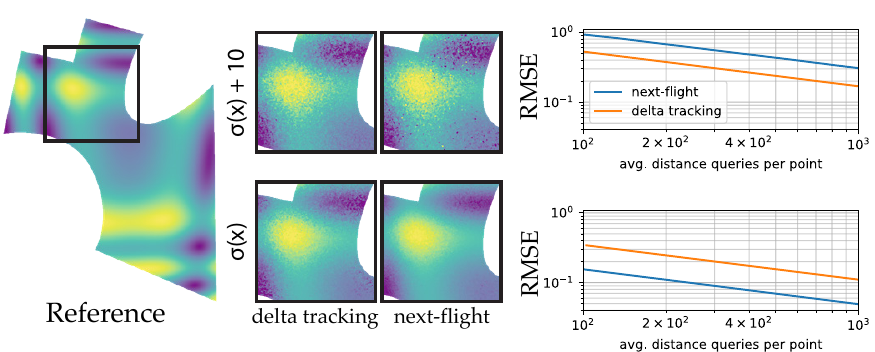}\\
    \caption{As with volume rendering, the effectiveness of our WoS algorithms varies based on the profile of the input coefficients. Here we show solution estimates computed with the delta tracking and next-flight variants on two reference analytic problems of the form $\Delta u - \zo({x}) u = -\src({x})$ with the same solution but different absorption coefficients $\zo({x})$ and $\zo({x}) + 10$ from \cref{fig:nf-walklength}. \emph{Top Row}: The delta tracking estimator demonstrates lower variance than the next-flight estimator due to higher sample correlation in the latter. \emph{Bottom Row}: Large values of $\zo({x})$ (and hence $\uCtrl$) cause the efficiency of the delta tracking estimator to drop, since more distance queries are needed to achieve the same variance as the next-flight estimator.}%
    \label{fig:nf-convergence}
\end{figure}

\begin{figure*}[t]
    \footnotesize
    \centering
    \includegraphics[width=\textwidth]{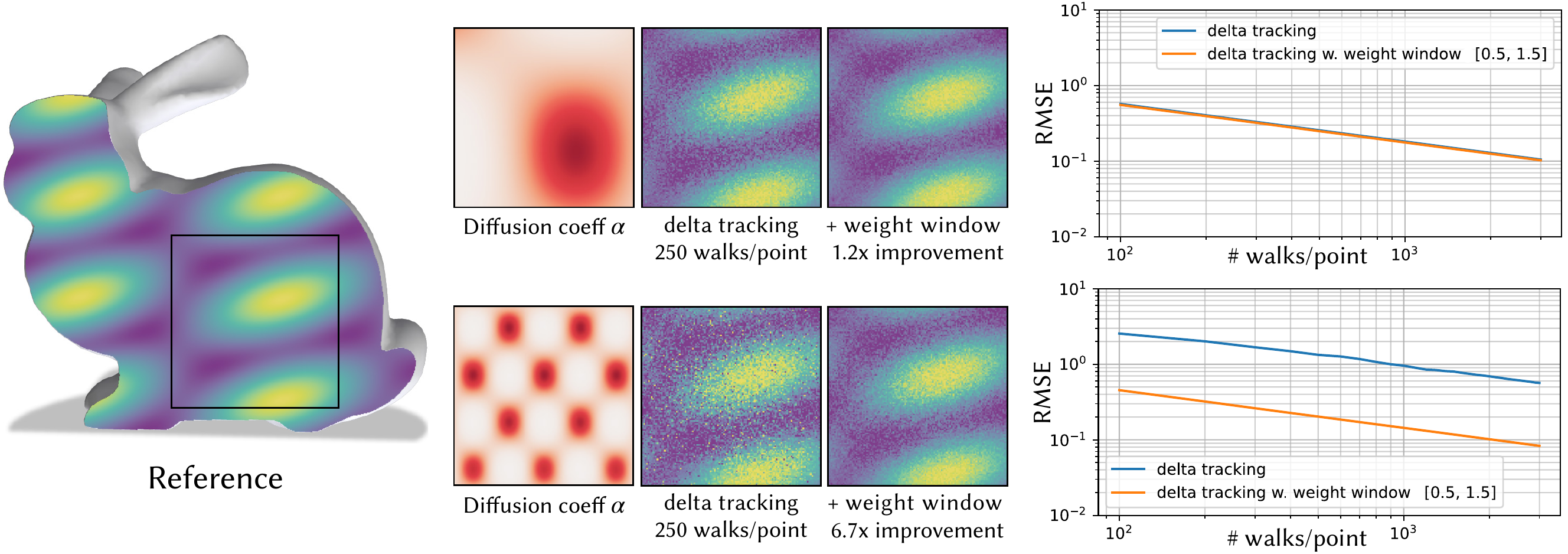}
  \caption{A static weight window is effective in reducing the variance of solution estimates computed with our WoS algorithms, as shown here on two reference analytic problems of the form $\nabla \cdot (\so({x}) \nabla u) = -\src({x})$ with the same solution but different diffusion coefficients $\so({x})$. A weight window provides significant variance reduction in problems involving high frequency diffusion coefficients (\emph{Bottom Row}). While a weight window is not necessarily required for low frequency coefficients (\emph{Top Row}), it often has the additional benefit of improving run-time performance by terminating walks using Russian Roulette.}
  \label{fig:solution-convergence-ww}%
\end{figure*}

The next-flight variant of WoS does not require any distance queries to evaluate $\est{\pathtr}$ and $\est{\pathsrc}$ inside $\ball({x}_k)$. This performance benefit does however come at the cost of increased correlation in $\est{\pathtr}$ due to the reused value of $\est{u}({x}_{k+1})$; see \cref{fig:nf-convergence}. In terms of parameters, the choice of how to pick $\uCtrl$, $\pdfBall$ and $\pdfSurf$ remains unchanged from \cref{sec:wos-delta-tracking}. Notice though that the Green's functions $G^{\uCtrl}({x}^l, {x}^{l+1})$ and Poisson kernels $P^{\uCtrl}({x}^l, {z})$ in this algorithm are no longer evaluated at the center of each ball $\ball({x}_k)$ for $l > 0$, but rather at arbitrary points inside $\ball({x}_k)$. Sec. $1.2$ and Sec. $1.4$ of the supplemental material provide expressions for evaluating and sampling \emph{off-centered} versions of these quantities.

\section{Variance Reduction}\label{sec:variance-reduction}

As shown in \cref{fig:nf-convergence}, \cref{fig:solution-convergence-ww}, \cref{fig:solvers-convergence} and \cref{fig:discrete-bias}, our WoS algorithms demonstrate the expected Monte Carlo rate of convergence when estimating the solution to variable-coefficient PDEs. Here we present a variance reduction strategy to reduce noise in the PDE estimates.

\paragraph{Weight Window.} Our algorithms use the Monte Carlo estimator in \cref{eq:pde-estimator} to solve variable-coefficient PDEs. For every step in a walk, the value of the estimate $\est{u}({x}_k)$ is scaled by a multiplicative weighting factor (with a value possibly greater than $1$). For instance, in the delta tracking variant of WoS, $\est{u}$ is weighted either by $\left(1 - \frac{\zoh({y}_{k+1})}{\uCtrl}\right)\sqrt{\frac{\so({y}_{k+1})}{\so({x}_k)}}$ or $\sqrt{\frac{\so(x_{k+1})}{\so({x}_k)}}$ based on whether the volume or the boundary term is sampled (resp.). These weights serve as a source of variance in the estimator; see \cref{fig:solution-convergence-ww}. This is especially true for large values of the bounding parameter $\uCtrl$, since a high average number of steps per walk can result in a cumulative weight that is either very small or very large. To address this issue, we recommend using a \emph{weight window}, a commonly employed variance reduction tool in neutron transport \citep{Hoogenboom:2005:Critical}, to keep the weighting factor roughly constant for any walk.

\begin{wrapfigure}[7]{r}{0.3\columnwidth}
\vspace*{-1\baselineskip}\hspace*{-0.06\columnwidth}\includegraphics{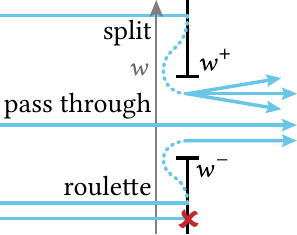}
\end{wrapfigure}
A weight window achieves variance reduction though a combination of \emph{Russian Roulette} and \emph{splitting} (see inset). The former terminates walks with small weights, since it is often not worthwhile completing walks whose likely contribution to the estimator is small. The latter splits walks with large weights into multiple new equally weighted walks. This prevents any single walk from having a large weight, while encouraging better exploration of the domain through the newly spawned walks. We adopt a simple version of the weight window strategy that uses a static window size $[\wwLo, \wwHi]$, where $\wwLo$ and $\wwHi$ are the minimum and maximum cumulative weight (resp.) any walk is allowed to have. In our implementation, we use $[0.5, 1.5]$. A walk whose cumulative weighting factor $w$ at any step lies within the window is allowed to continue without modification. A Russian Roulette survival probability of $w/\wwLo$ determines whether to terminate the walk when $w < \wwLo$. Otherwise, the walk is split into $m \coloneqq w/\wwHi$ new walks, each with a weight of $w/m$ when $w > \wwHi$. Since $m$ is generally not an integer, we use the \emph{expected value splits} approach of \citet{Booth:1985:Split} where $n \coloneqq \lfloor m \rfloor$ walks are chosen with probability $n + 1 − m$, and $n + 1$ walks are chosen otherwise.

\cref{fig:solution-convergence-ww} highlights the effectiveness of a static weight window in reducing variance in problems with high frequency coefficients and large bounding parameters $\uCtrl$. In the neutron transport and rendering literature, specifying the window size adaptively has been shown to provide improvements in computational efficiency by up to an order of magnitude over static windows \citep{Booth:1984:Importance, Wagner:1998:Automated, Vorba:2016:Adjointdriven}. We leave this optimization to future work.

\section{Implementation and Results}\label{sec:framework}

In this section, we discuss practical considerations pertaining to the modified WoS algorithms we present. We also highlight the unique benefits of our approach on a range of example problems inspired by engineering and design applications.

\subsection{Algorithm Inputs}\label{sec:algorithm-inputs}

Walk on spheres requires computing the distance to the boundary of a domain to find the largest empty sphere in each step of a random walk. Distance queries can be accelerated via a spatial hierarchy such as a bounding volume hierarchy (BVH) or an octree \citep{Ericson:2004:Collision}. These acceleration structures can be built quickly and without much memory across a large variety of geometric representations of the domain boundary, e.g., polygonal meshes, NURBS/subdivision surfaces, implicit surfaces and constructive solid geometry \citep{TheEmbreedevelopers:2013:Embree, Museth:2013:VDB}. Our CPU implementation currently uses a basic axis-aligned BVH without vectorization, though recent GPUs provide opportunities for further acceleration \citep{Burgess:2020:Rtx}.

Our WoS algorithms for variable-coefficient PDEs additionally require computing the gradient and Laplacian of the diffusion and transport coefficients $\so({x})$ and $\fo({x})$ (see \cref{sec:transformations} and \cref{app:stochastic-pde-derivation}).  These derivatives can be computed using any standard technique (e.g., auto-differentiation).  Like most volume rendering algorithms \citep{Novak:2018:Monte}, we also require computing a bounding parameter $\uCtrl \coloneqq \max(\zoh({x})) - \min(\zoh({x}))$ over the domain.



\subsection{Geometric Flexibility and Iterative Feedback}\label{sec:geometric-flex-feedback}

Monte Carlo methods are popular in rendering because they provide instantaneous feedback and easily work with a wide variety of geometric representations. This enables engineers, scientists and artists to quickly iterate on their designs, however complex they maybe. The same is true for the solvers we present in this paper.

\begin{figure}[t]
    \centering
    \footnotesize
    \includegraphics[width=\columnwidth]{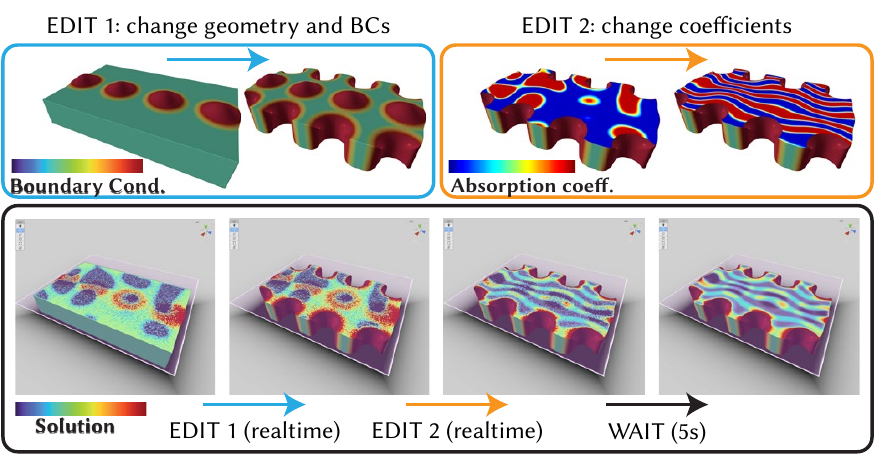}%
    \caption{Our method is ideal for interactive editing applications since we can operate directly on edit-time scene representations (here a list of signed distance field CSG operations) and refine the solution iteratively when updating the scene (EDIT 1) or the coefficients (EDIT 2). This provides designers with immediate feedback which can greatly reduce iteration time.}
    \label{fig:interactive-editing}
\end{figure}

We implemented a GPU version of our solver in the Unity game engine, allowing a user to work with variable-coefficient PDEs interactively (\cref{fig:interactive-editing}), even on scenes of immense complexity (\cref{fig:teaser}). We represent the domain geometry using signed distance fields and can use this same representation to interactively edit the scene geometry, render it using sphere tracing \cite{Hart:1996:Sphere}, and solve a PDE using our WoS estimators---all without resorting to any meshing.

\Cref{fig:teaser} shows a collection of objects scattered procedurally to construct an infinitely large yet aperiodic scene. Meshing this entire scene is not possible since it requires infinite storage. In fact, even meshing a subset of the scene is prohibitively expensive (see \cref{fig:amr}). Since our method operates directly on the signed distance field and supports local point evaluation, we can compute the solution in only the region visible to the camera and resolve high frequency coefficients despite the large scale of the scene.

In \cref{fig:interactive-editing} we show how a user can interactively edit the scene geometry via constructive solid geometry operations while seeing the solution to the variable coefficient PDE on a slice plane.
We perform a single random walk for each point on the slice plane each time we render a frame.
When the geometry or coefficients change, we discard prior samples, otherwise we accumulate samples and average.
Our solver quickly produces a converged solution, since it runs at 60 frames per second, meaning that it can produce 60 samples per point per second.
As a result, users get instant feedback when they change problem inputs, enabling them to guide the design process effectively without any interruptions.

\subsection{Integration with other Monte Carlo solvers}\label{sec:subsurface-scattering}

One benefit of Monte Carlo solvers is that point evaluation makes it trivial to combine solvers to estimate the solution to nested or coupled problems.
As an example of this, we show how to use the Monte Carlo estimators presented in this paper to accelerate volumetric rendering.

The VRE we presented in \cref{sec:volume-rendering} is already solved almost exclusively using Monte Carlo methods \cite{Novak:2018:Monte}, at least in the context of offline rendering.
Unfortunately, this is expensive for dense, high-albedo media like milk or marble, where light paths can undergo thousands of scattering events.
In these cases, it is common to use a diffusion equation like \cref{eq:pde} to model light transport, because it is much faster~\cite{Jensen:2001:Practical}.  The resulting 3D heterogeneous diffusion problem can be computed using classical deterministic solvers~\citep{Arbree:2011:Heterogeneous, Koerner:2014:Fluxlimited}, but these methods have seen limited use in rendering due to their need for meshing, global solves, and great care in coupling the diffusion solver with the overarching Monte Carlo rendering algorithm.
Practical approaches have typically relied on approximations like homogeneity and local planarity via the dipole and its variants~\citep{Jensen:2001:Practical,Donner:2008:Layered,dEon:2011:Quantizeddiffusion,Habel:2013:Photon}---which typically introduce more error than the diffusion assumption itself. Our approach instead allows us to couple a Monte Carlo VRE solver with a fully heterogeneous 3D diffusion solver while avoiding many of these compromises.

Our hybrid algorithm uses a simple heuristic to switch on the fly between solving the RTE via classical volumetric path tracing (VPT) and solving a diffusion equation using our WoS estimators. At each path vertex inside the medium, we generate a free-flight distance (as in VPT) \emph{and} compute the distance to the medium boundary (as in WoS). We then continue the path with VPT or WoS based on which distance is larger. A path can freely change back and forth between these two modes, but in practice the algorithm will prefer the more accurate VPT steps near the boundary, while large WoS steps allow the algorithm to make rapid progress through the interior of the medium. This is conceptually very similar to shell tracing \cite{Moon:2008:Efficient, Muller:2016:Efficient} in rendering and condensed history neutron transport \cite{Fleck:1984:Random}: each WoS step essentially aggregates an arbitrary number of VPT steps.

\begin{figure}[t]
    \centering
    \includegraphics{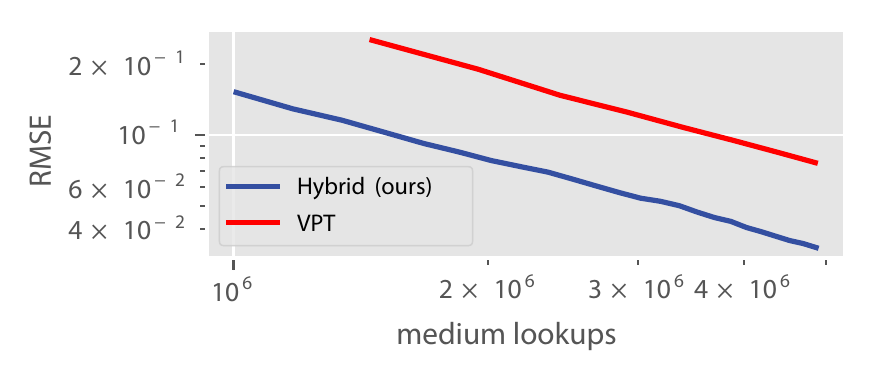}\\
    \includegraphics{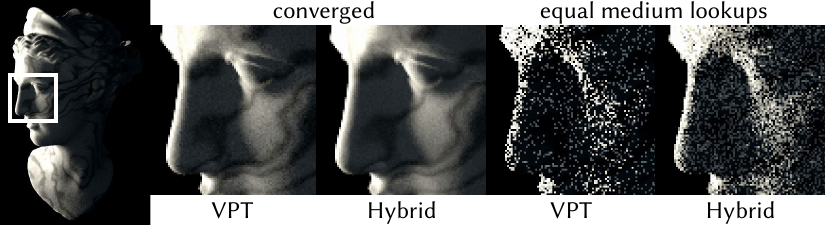}\\
    \caption{\emph{Top:} Our hybrid method demonstrates consistently lower relative mean squared error (RMSE) at equal number of medium lookups compared to VPT. \emph{Bottom:} Visual comparisons of our hybrid method and VPT.}%
    \label{fig:vpt-relMSE}
\end{figure}

In \cref{fig:vpt-relMSE} we compare this hybrid method to VPT on a heterogeneous marble bust, showing that even this simple proof-of-concept heuristic can achieve significant error reduction at equal medium lookups. Note that unlike techniques based on the dipole approximation, our diffusion solver accounts for the 3D heterogeneity inside the medium.

\subsection{Heterogeneous Diffusion Curves}
\label{sec:HeterogeneousDiffusionCurves}

\begin{figure}[t!]
    \centering
    \includegraphics{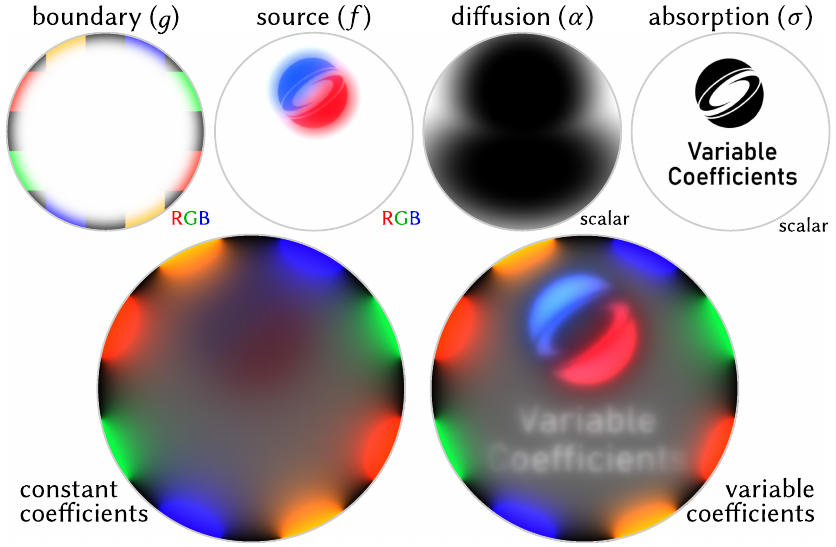}%
    \caption{Our method enables us to augment classic diffusion curve images with spatially-varying coefficients that provide greater artistic control.  For instance, source terms $\src$ have limited use in classic diffusion curves, since they get severely blurred out \emph{(left)}. By locally adjusting the diffusion strength $\so$, we can decorate the image with interesting textures or decals \emph{(right)}.}
    \label{fig:HeterogeneousDiffusionCurves}
\end{figure}

WoS is attractive for \emph{diffusion curves}~\cite{Orzan:2008:Diffusion}, since it provides real-time progressive previews that are easily implemented on the GPU~\cite{Quilez:2020:MCP}, can be applied to a zoom-in without first computing a coarse global solution (as done by \citet[Section 3.2.4]{Orzan:2008:Diffusion}), and avoids aliasing of fine features due to grid resolution~\cite[Figure 16]{Sawhney:2020:Monte}.  Our method makes it possible to generalize classic diffusion curves by also painting a source $\src(x)$, diffusion coefficient $\so(x)$, and absorption coefficient $\zo(x)$---\cref{fig:HeterogeneousDiffusionCurves} shows one example.  The added benefit is that, unlike constant-coefficient diffusion images, the source term $\src(x)$ is no longer severely blurred---enabling one to add interesting decals or background texture, while still smoothly diffusing color over regions without such details. Variable absorption $\zo(x)$ helps to further emphasize detail, since the strength of the source contribution is roughly $1/\zo(x)$. Akin to ``texture shaders''~\citep{Bowers:2011:Ray,Prevost:2015:Vectorial}, this enables an enriched design space spanning diffusion curves and traditional 2D graphics, though these prior approaches achieved this via alpha blending.

\subsection{Walk on Curved Surfaces}
\label{sec:WalksOnCurvedSurfaces}

A basic hypothesis of the original WoS algorithm is that a random walk exits every point on the boundary of a ball with equal probability (\cref{sec:walk-on-spheres-algorithm}).  However, on surfaces with non-constant curvature this hypothesis no longer holds: intuitively, more walkers will escape through the ``valleys'' than through the ``mountains.''  As a result, standard WoS cannot be used for many algorithms in geometric and scientific computing that need to solve equations on a surface/shell.

\setlength{\columnsep}{1em}
\setlength{\intextsep}{0em}
\begin{wrapfigure}{r}{137pt}
    \centering
    \includegraphics{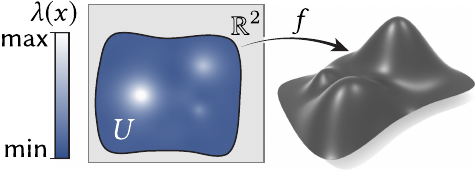}
\end{wrapfigure}
Our variable-coefficient scheme enables WoS to be applied to curved surfaces for the first time.  In particular, consider any surface expressed as a \emph{conformal parameterization} \(f: \mathbb{R}^2 \supset U \to \mathbb{R}^3\); conformal means that \(f\) distorts the surface by a uniform scaling \(\lambda({x})\) at each point \({x} \in U\), i.e., \(J_f^T J_f = \lambda(x)I\), where \(I\) is the identity and \(J_f\) is the Jacobian of \(f\).  The Laplacian \(\Delta_f\) of the curved surface is then related to the ordinary Laplacian via \(\Delta = \lambda\Delta_f\).  Hence, we can solve PDEs on the curved surface by replacing the usual diffusion coefficient \(\alpha(x)\) with \(\lambda(x)\lambda(x)\).  \cref{fig:surface-examples} shows several examples; for periodic domains (like the torus) our walks simply ``wrap around.''  In theory, \emph{every} surface admits a conformal parameterization (by the uniformization theorem~\cite{Abikoff:1981:UT}), but in practice many important surfaces used in engineering (such as NURBS or other spline patches) are expressed in non-conformal coordinates.  To directly handle such patches, we would need to extend our method to \emph{anisotropic} diffusion coefficients---an important topic for future work.  Note also that earlier work on diffusion curves for surfaces uses free-space 2D Green's functions~\cite{Sun:2012:Diffusion}, which in general provide only a rough proxy for the true Green's functions of the curved surface.

\begin{figure}
   \includegraphics[width=\columnwidth]{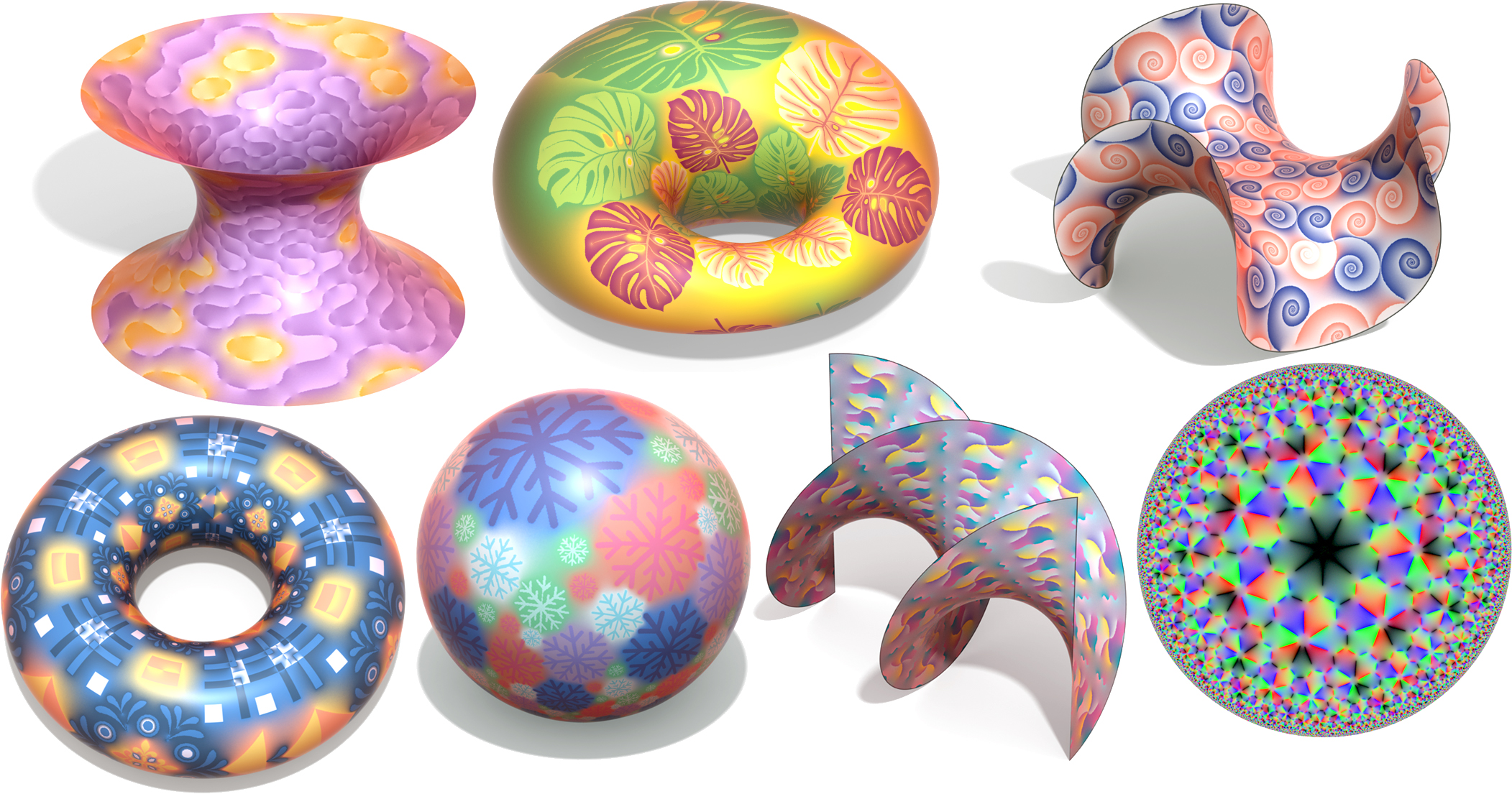}
   \caption{Variable coefficients enable us to extend WoS to curved domains (not previously possible).  Here we directly resolve intricate boundary conditions for diffusion curves---without generating a fine surface mesh that conforms to boundary curves, nor a spatially-adaptive grid in the parameter domain.\label{fig:surface-examples}}
\end{figure}

\section{Related Work and Comparisons}
\label{sec:RelatedWorkAndComparisons}

\subsection{PDE Solvers}
\label{sec:PDESolvers}

There are innumerable methods for solving variable-coefficient PDEs---here we examine core trade offs between major classes of methods, highlighting unique benefits of Monte Carlo that complement existing solvers.  Note that some methods \emph{homogenize} coefficients to a coarser model that approximates fine-scale behavior \citep{Efendiev:2009:Multiscale, Abdulle:2012:Multiscale, Durlofsky:1991:Blockperm, Dykaar:1992:Conductivity, March:2021:Heterogeneous}. We instead seek to directly resolve fine details---in fact, Monte Carlo methods like ours may help to accurately compute parameters for homogenization.

\subsubsection{Finite Element Methods (FEM)}
\label{sec:FiniteElementMethods}

All finite element methods (including meshless and boundary element methods) adopt a common mathematical framework: given a linear PDE \(Lu = f\), find the best approximation to \(u\) within a finite-dimensional function space.  E.g., standard \emph{Galerkin FEM} considers an approximation \(\widehat{u} := \sum_{i=1}^n u_i \phi_i\) in some basis \(\phi_1, \ldots, \phi_n: \Omega \to \mathbb{R}\), where the \(u_i \in \mathbb{R}\) are unknown coefficients.  Letting \(\langle u, v \rangle \coloneqq \int_\Omega u(x) v(x) \diff x\) denote the \(L^2\) inner product, one then seeks a \(\widehat{u}\) satisfying
\begin{equation}
   \label{eq:WeakFormulation}
   \langle L\widehat{u}, \phi_j \rangle = \langle f, \phi_j \rangle, \quad j = 1, \ldots, n,
\end{equation}
\ie, such that \(\widehat{u}\) looks exactly like the true solution \(u\) when restricted to the subspace \(V := \text{span}(\{\phi_i\})\) (but may have error in directions orthogonal to \(V\)).  To solve this equation, we can rewrite \cref{eq:WeakFormulation} as
\[
   \textstyle\sum_{i=1}^n u_i \langle L \phi_i, \phi_j \rangle = \textstyle\sum_{i=1}^n f_i \langle \phi_i, \phi_j \rangle.
\]
The inner products between basis functions and their derivatives define elements of the \emph{mass} and \emph{stiffness} matrices (resp.), and must typically be approximated via numerical quadrature.

\begin{figure}[t]
    \centering
    \includegraphics[width=\columnwidth]{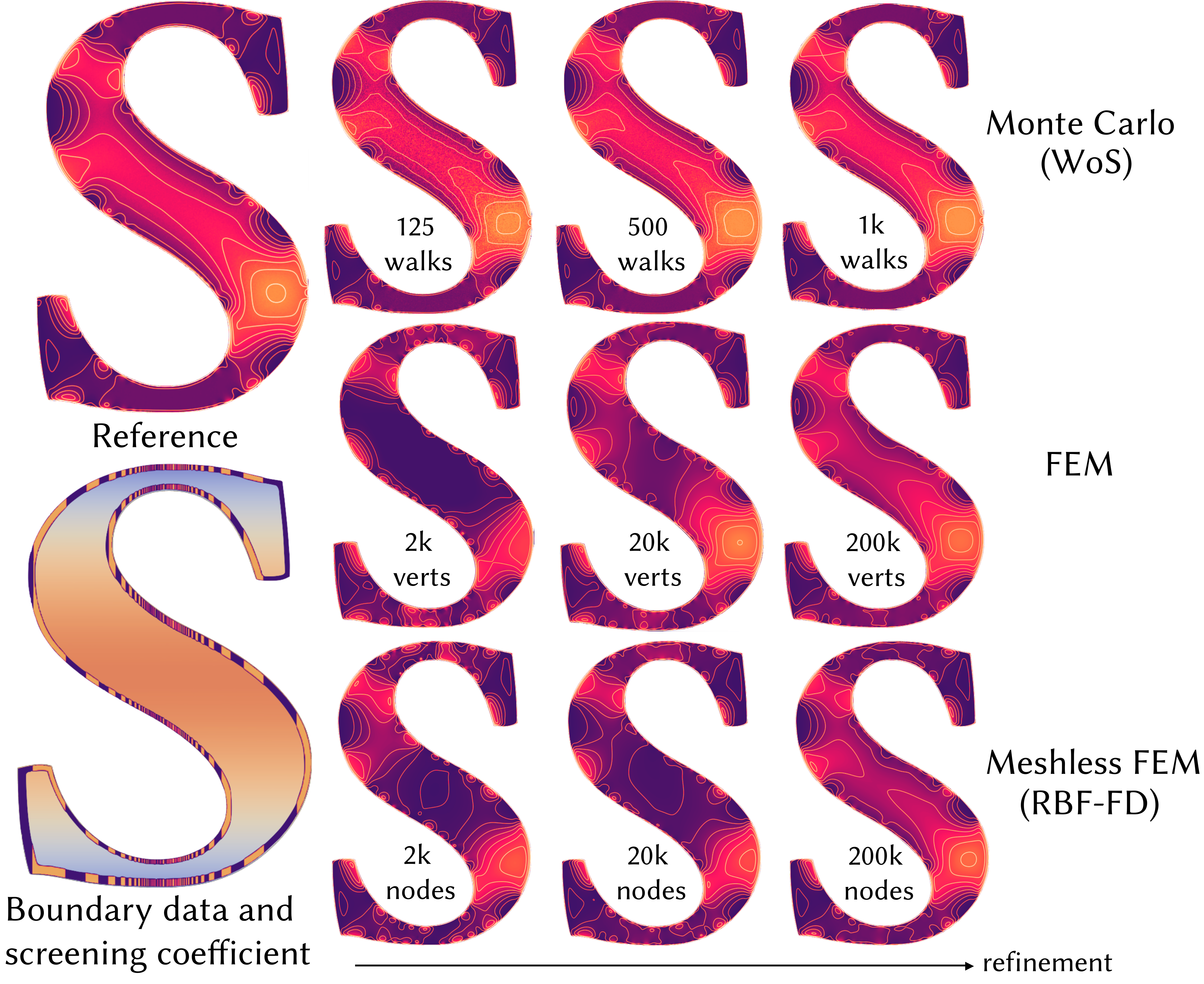}
    \caption{With both FEM and MFEM, local aliasing of high-frequency boundary data yields large global errors in the solution, demanding significant refinement.  In contrast, WoS always captures the global solution (even with very few samples); error instead manifests as high-frequency noise.\label{fig:spatial-aliasing}}
\end{figure}

From this perspective, the only difference between flavors of FEM is the choice of basis functions \(\phi_i\). Hence, all finite element methods come with a common set of challenges:
\begin{itemize}
   \item They must all solve a globally-coupled system of equations.
   \item They are all prone to spatial aliasing in the geometry, solution, boundary conditions, source terms, and/or coefficients, since any finite basis \(\{\phi_i\}\) provides limited spatial resolution.
   \item They all demand spatial discretization (meshing or sampling) to define bases \(\phi_i\), which can be costly and error prone (\cref{fig:discretization-challenges}).
\end{itemize}

In contrast, WoS can directly evaluate the solution at any point without meshing/sampling, and without performing a global solve.  Moreover, it does not suffer from aliasing in the solution or problem data, since it need not restrict functions to a finite-dimensional subspace (see especially \cref{fig:spatial-aliasing}).

\begin{figure}[t]
    \centering
    \includegraphics{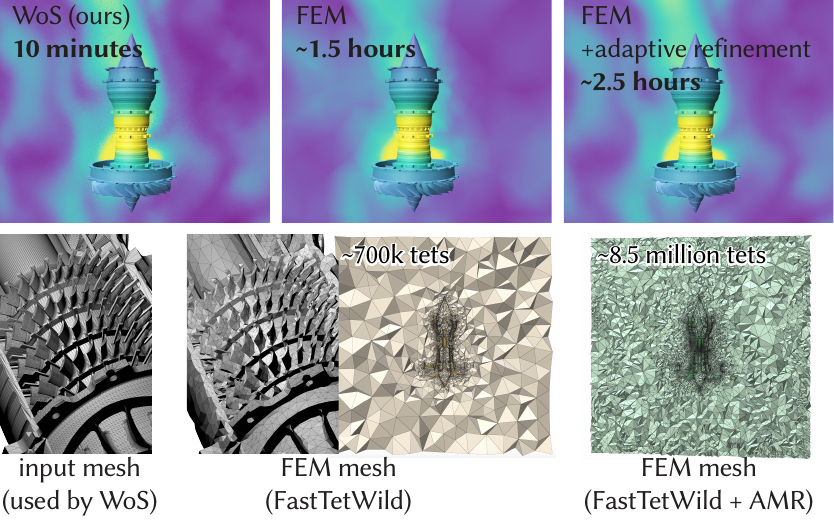}
    \caption{Even when starting with a coarse approximation of just a single black body from \cref{fig:teaser}, FEM takes immense time and memory to resolve detailed variations due to PDE coefficients.  Here, initial coarse meshing by \emph{FastTetWild}~\cite{Hu:2020:fTetWild} takes about 1.5 hours to produce a mesh that cannot resolve fine details in solution or geometry (despite being quite large already). After ~1 hour more of AMR via \cite{Anderson:2021:MFEM} the solution is better resolved, but the mesh size has blown up.  Our CPU-based WoS implementation takes about 10 minutes total, on the same machine.}
    \label{fig:amr}
\end{figure}

\paragraph{Mesh-Based FEM} Most often, FEM bases \(\phi_i\) are defined via polyhedral mesh elements.  Quickly and robustly meshing large, detailed and/or imperfect geometry (e.g., with self-intersections) is an ongoing ``grand challenge,'' where even state-of-the-art methods can struggle (\cref{fig:discretization-challenges}).  This problem gets harder if the mesh must also be refined for spatially-varying coefficients: even with intelligent \emph{adaptive mesh refinement (AMR)} \citep{Zienkiewicz:1992:Superconvergent1, Zienkiewicz:1992:Superconvergent2}, meshing quickly becomes prohibitive (\cref{fig:amr}).  More recent \emph{a priori} \(p\)-refinement does not help, since it considers only element quality and not spatial frequencies in the solution or problem data \citep{Schneider:2018:Decoupling}. WoS bypasses the meshing grand challenge entirely, needing only a BVH (for closest point queries), which uses minimal memory and can be built in a fraction of a second---even for totally degenerate geometry (see \cite[Figure 2]{Sawhney:2020:Monte}).

\begin{figure}[t]
    \centering
    \includegraphics[width=\columnwidth]{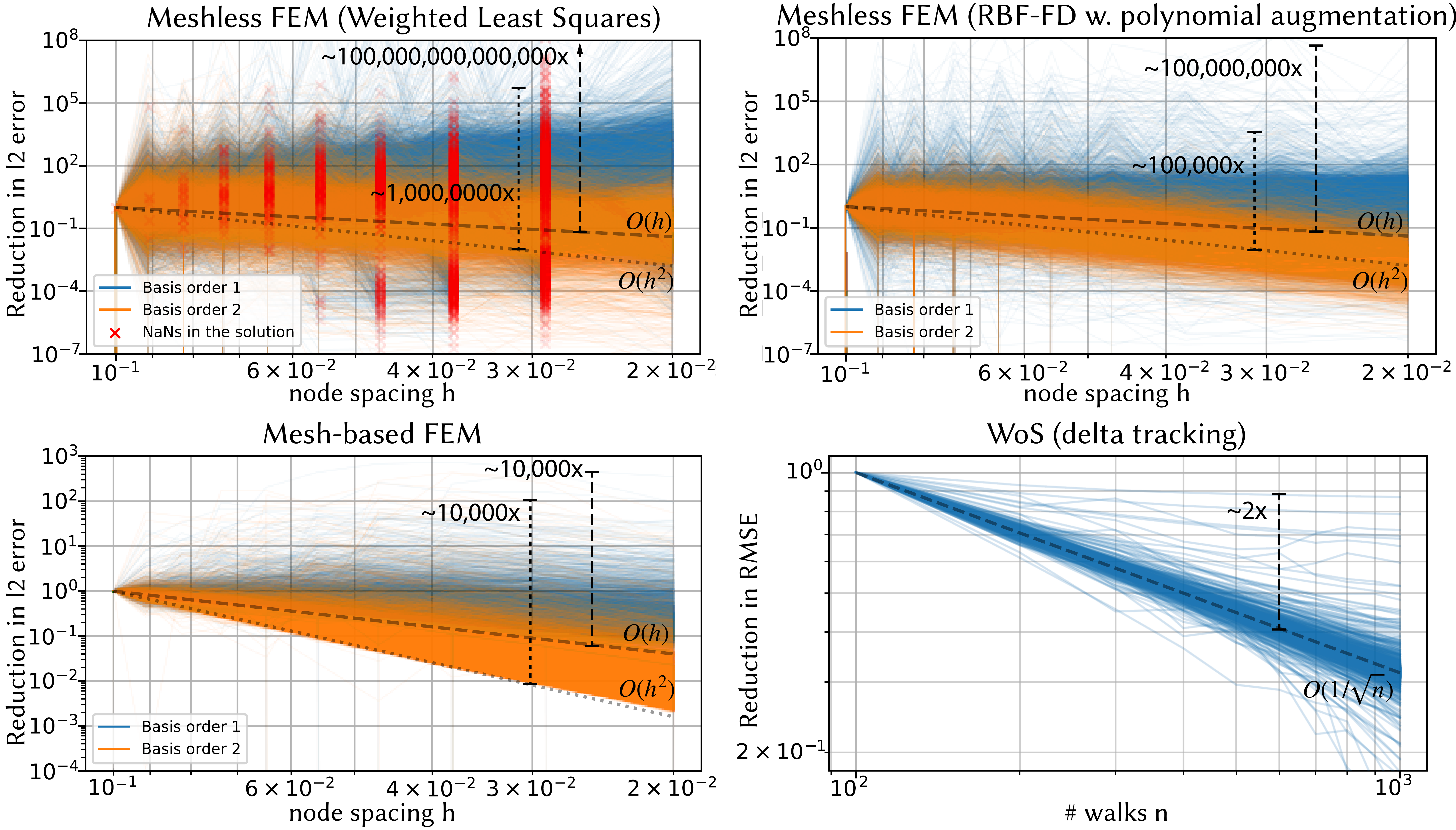}
    \caption{Convergence criteria for meshless FEM methods are not well-understood. In practice, methods often fail to converge under refinement (shown here for the same variable-coefficient problem on watertight models generated from the Thingi10k dataset \citep{Thingi10K, Hu:2020:fTetWild}) and/or show extremely large variation in error. Both mesh-based FEM and Monte Carlo methods come with mathematically rigorous convergence guarantees and behave much more predictably under refinement.}
    \label{fig:solvers-convergence}
\end{figure}

\paragraph{Meshless FEM} Though \emph{meshless FEM (MFEM)} may sound like a natural alternative to grid-free Monte Carlo, the term ``meshless'' is a bit of misnomer: MFEM does not need a polyhedral mesh, but must still discretize the domain by carefully arranging nodes on the domain interior (\cref{fig:MeshlessMeshing}).  Nodes are then associated with bases \(\phi_i\), such as a \emph{radial basis functions (RBFs)}, to build mass and stiffness matrices. To couple bases with overlapping support one must identify neighbors---forming a global graph structure similar in size and complexity to a polyhedral mesh.  Node locations must satisfy criteria that are often just as difficult and delicate to enforce as mesh quality criteria \citep[Ch. 3]{Li:2007:Meshfree}, and often require global optimization akin to mesh smoothing \citep{Slak:2019:Nodegen}. Moreover, just as one bad element can ruin an FEM solution, bad node placement can leading to catastrophic failure (e.g., NaNs in the solution---see \cref{fig:solvers-convergence}, \emph{top left}). To mitigate spatial aliasing one can adaptively sample nodes---but unlike mesh-based FEM, adaptive refinement for MFEM is poorly understood (lacking, e.g., rigorous convergence guarantees).  Finally, typical MFEM bases are \emph{approximating} rather than \emph{interpolating}, complicating the enforcement of boundary conditions \citep{Fries:2004:CMF, Nguyen:2008:MFEMreview}; some MFEM methods hence modulate bases by a distance-like function \cite{Shapiro:1999:MSD}, but still effectively spatially discretize functions on the interior by choosing a finite basis \(\{\phi_i\}\).

\begin{figure}[t]
    \centering
    \includegraphics[width=\columnwidth]{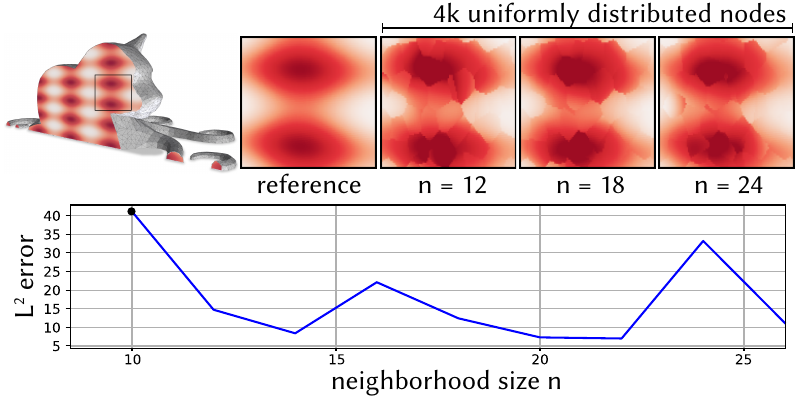}
    \caption{In practice, it can be difficult to find reliable parameters for meshless FEM---for instance, increasing neighborhood size often \emph{increases} error in an unpredictable way.  In contrast, WoS requires no parameter tuning.\label{fig:meshless-neighborhood}}
\end{figure}

As shown in \cref{fig:solvers-convergence}, a more serious challenge with meshless FEM is \emph{stagnation}: until only very recently \citep{Flyer:2016:RBF1, Bayona:2017:RBF2, Bayona:2019:RBF3}, MFEM methods might fail to converge without careful problem-specific tuning of parameters such as neighborhood size. More damning is the fact that increasing the neighborhood size does not always make the solution better (see \cref{fig:meshless-neighborhood}).  Moreover, whereas convergence of \emph{adaptive} FEM is rigorously understood~\cite{Mekchay:2005:CAF}, there is a dearth of corresponding results for adaptive MFEM schemes---especially important for problems with fine details in geometry and/or coefficient functions.  In practice, MFEM also requires far denser mass/stiffness matrices than those used in mesh-based FEM, while often providing less-accurate results. For instance, methods such as \emph{RBF-FD with polynomial augmentation} \citep{Flyer:2016:RBF1} that converge under refinement require bases of order at least 2. On the whole, MFEM is not known for its reliability---in stark contrast, walk on spheres guarantees that the expected solution is equal to the true solution of the smooth PDE, without any parameter tuning whatsoever. Moreover, unlike MFEM, WoS is truly ``meshless'': at no point in the algorithm does one require a global sampling or meshing of the domain.

\begin{figure}[t]
    \includegraphics{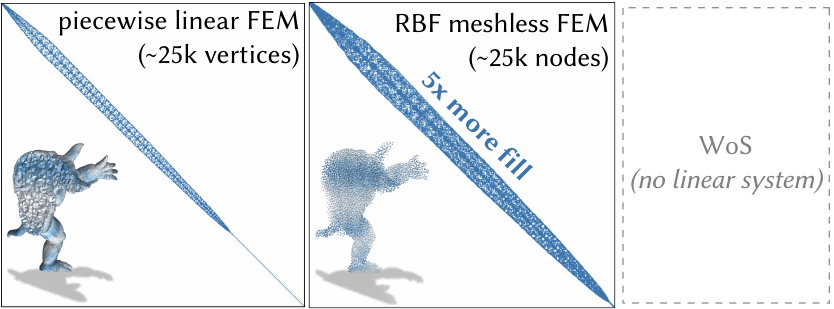}
    \caption{Meshless FEM must solve a much denser linear system than even standard FEM---whereas WoS avoids solving a global system altogether.\label{fig:fem-matrix-fill}}
\end{figure}

Finally, though meshless FEM has been around for a long time, it has not seen nearly as much use in practice as mesh-based methods (e.g., with very few open source or commercial packages available relative to mesh-based FEM). \All{TODO This needs to be fixed; more like: there are commercial MFEM packages, but have not been widely adopted despite being around for a long time.}

\begin{figure}[t]
    \centering
    \includegraphics[width=\columnwidth]{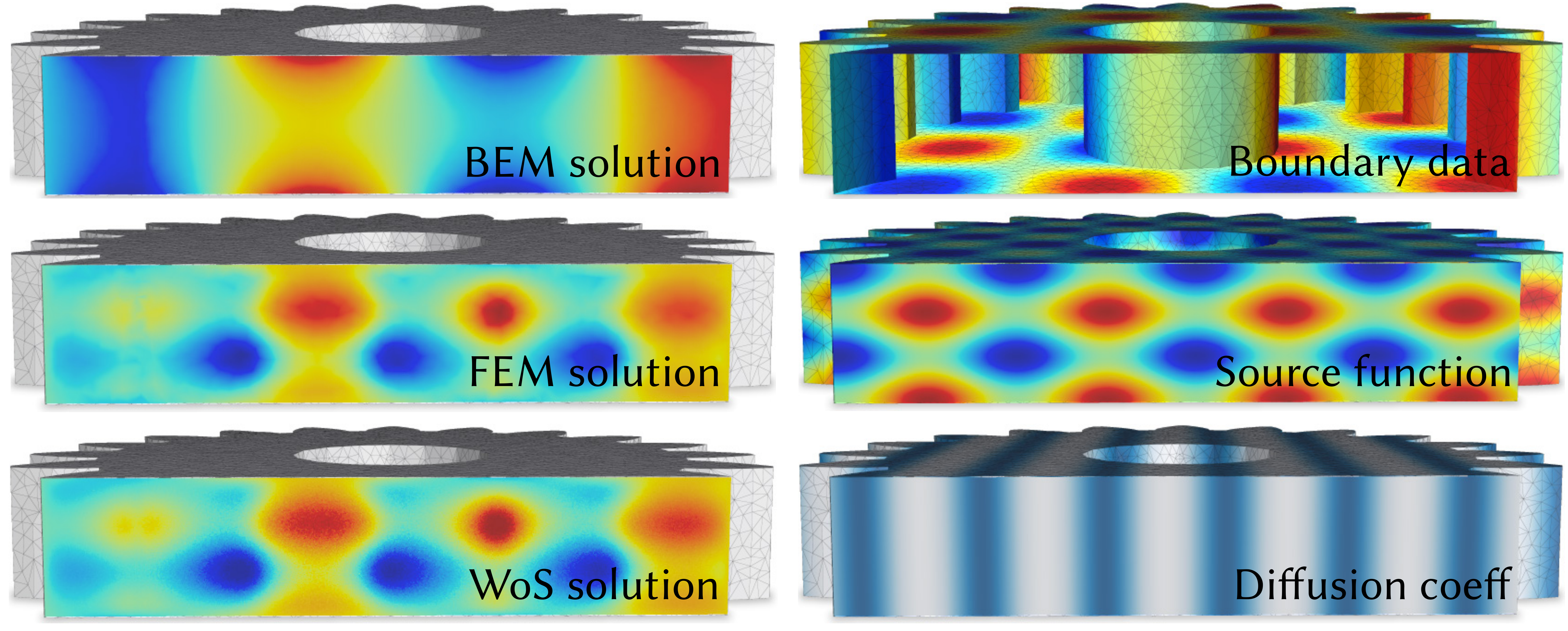}
    \caption{Unlike FEM and Monte Carlo, traditional BEM ignores volumetric functions (e.g. coefficients and source terms) that affect the PDE solution.}
    \vspace{-1\baselineskip}
    \label{fig:bem}
\end{figure}

\paragraph{Boundary Element Methods (BEM)} Boundary element methods approximate the solution using bases functions \(\phi_i\) associated only with elements of a boundary mesh (such as free-space Green's functions). These methods draw a natural comparison with walk on spheres, since they need not discretize the interior of the domain.  However, there is a significant difference in capabilities: whereas WoS easily handles problems with source terms and spatially-varying coefficients on the domain interior, basic BEM ignores these terms altogether (see \cref{fig:bem}). In order to handle general interior terms, one must couple BEM with a second, interior solver such as FEM, MFEM, or FD---inheriting all the same challenges~\citep{Partridge:2012:DRBEM,Costabel:1987:BEM,Coleman:1991:EBE}. Moreover, even for problems involving only boundary terms, BEM must discretize the boundary geometry, leading to spatial aliasing in both boundary data and geometry. Unlike FEM/MFEM, BEM must solve a globally-coupled \emph{dense} system of equations, demanding special techniques like \emph{hierarchical matrix approximation}~\citep{Hackbusch:2015:HMA} to obtain reasonable performance).

\subsubsection{Finite Difference (FD) Methods}
\label{sec:FiniteDifferenceMethods}

The main conceptual difference that distinguishes all finite difference methods from finite element methods is that the degrees of freedom are no longer coefficients in a finite basis---instead, they represent point samples of the unknown function at nodal points. Derivatives are then approximated essentially via Taylor series approximation: if one imagines that there is a function interpolating the nodal values, then the derivatives are given in terms of standard finite difference formulas.

On the one hand, FD schemes are attractive due to the simplicity of implementation on a regular grid.  They are however less than ideal for PDEs with nonuniform coefficients, where the uniformity of grid cells can lead to significant numerical diffusion \citep{Umansky:2005:NSS}, spurious negative values \citep{Sharma:2007:PMA}, and locking/stagnation \citep{Babuska:1992:LRF}. Another major challenge is spatial adaptivity: for many PDEs, using a fine grid over the whole domain is overkill. In order to adaptively refine solutions (e.g., due to the impact of variable coefficients) one can switch to a hierarchical structure like an octree \citep{Losasso:2006:SLF, Gibou:2018:LSA}. Octrees mitigate some of the performance advantage of working with a regular grid, and significantly increase the complexity of implementation. FD methods also struggle with accurate handling of boundary conditions due to the fact that cell boundaries are now axis-aligned \citep{Causon:2010:FDPDE}.

On the whole, finite differences suffer from the same major challenges as finite element methods: one must spatially discretize the domain, the boundary conditions, the source term, and the coefficients of the operator, invariably leading to either aliasing or oversampling. Moreover, one must solve a globally-coupled system of equations over the entire domain, rather than concentrating computational effort only at points of interest (as in Monte Carlo).

\paragraph{Material Point Methods.} \emph{Material point methods}~\citep{Jiang:2016:MPM}, such as PIC~\citep{Harlow:1965:NCT}, FLIP~\citep{Brackbill:1986:FMA,Zhu:2005:ASF}, APIC~\citep{Jiang:2015:APM}, and MPM~\citep{Sulsky:1995:APM} are popular for time-dependent computational mechanics problems involving large scale deformation (fluids, plasticity, etc.). These methods are also sometimes referred to as ``meshless'', but they are not (in general) MFEM schemes as defined in \cref{sec:FiniteElementMethods}. Rather, these methods use particles to approximate advection, and a background grid to solve elliptic problems (such as pressure projection in fluids).  Critically, for the problems we consider here (time-\emph{independent} elliptic PDEs), there is no advection component, and MPM reduces to simply solving elliptic equations on a grid---with the same tradeoffs as discussed above.

\subsubsection{Stochastic Methods}
\label{sec:stochastic-methods}

Not all PDE solvers need to discretize space---the notable exception are Monte Carlo methods based on continuous random processes such as Brownian motion. The stochastic approach to deterministic boundary value problems, discussed in \cref{sec:background}, centers on the simulation of random walks that in aggregate solve a large class of 2nd-order elliptic PDEs \citep{Oksendal:2003:Stochastic}. This formulation enables local point evaluation of the PDE solution and has found extensive use in scientific disciplines such as mathematical finance \citep{Black:1973:OptionPricing, Merton:1971:Optimum, Merton:1992:ContinuousFinance, Cox:1985:Interest}, computational physics and chemistry \citep{Grebenkov:2007:NMR, Gillespie:1977:Chemical, Mascagni:2004:Molecules, Mascagni:2004:RWB} and optimal control \citep{Kalman:1960:filter, Kappen:2007:SCT} (albeit often on simple geometric domains).

\paragraph{Discretized Random Walks}

In place of WoS, one might try approximating Feynman-Kac by directly simulating the diffusion process \(X_t\) with explicit time stepping \citep{Kloeden:2013:NSS, Higham:2001:SDESim}, akin to \emph{ray marching}~\cite{Tuy:1984:DDO}:
\begin{equation}\label{eq:sde-discretized}
   X_{k+1} = X_k\ +\ \fo(X_k) h\ +\ \sqrt{\so(X_k)}\ (W_{k+1} - W_k).
\end{equation}
\begin{figure}[t]
    \centering
    \includegraphics[width=\columnwidth]{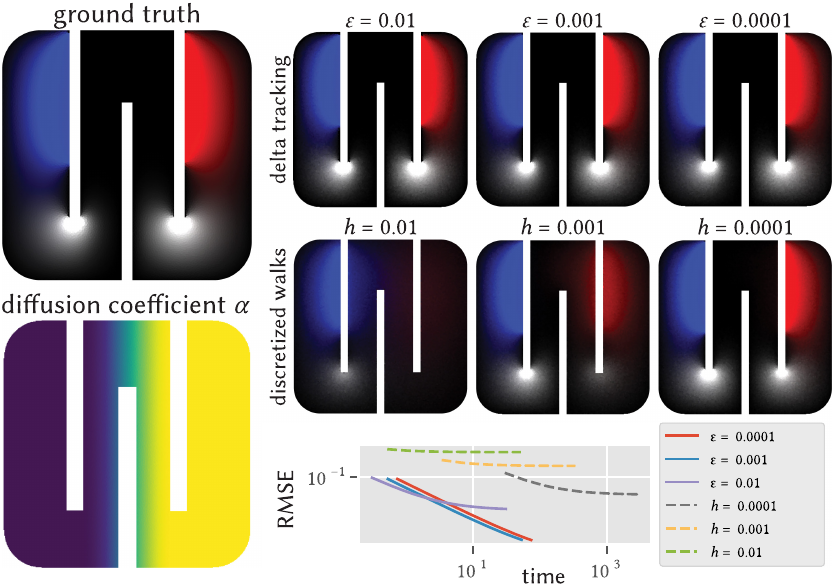}\\
    \caption{\emph{Top Row, Right}: Our WoS algorithms correctly resolve boundary conditions for any value of \(\varepsilon\); shrinking the \(\varepsilon\)-shell reduces the bias in the solution in a predicable manner, with little impact on performance (\emph{Bottom Right}). \emph{Middle Row, Right}: Solving the same PDE by simulating discretized random walks with the integration scheme in \cref{eq:sde-discretized} eventually resolves the boundary conditions with a finer step-size, though at the detriment of run-time performance (\emph{Bottom Right}).}%
    \label{fig:discrete-bias}
\end{figure}

\newcommand{\discreteSdeFigure}{
\setlength{\columnsep}{1em}
\setlength{\intextsep}{0em}
\begin{wrapfigure}{r}{1.2in}
    \centering
    \includegraphics{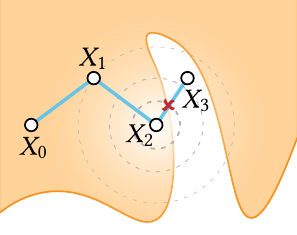}%
    \vspace*{-2ex}%
    \caption{Discretized random walks can leave the domain, biasing estimates.}
    \label{fig:sde-bias}
\end{wrapfigure}
}
However, this approach introduces several sources of error.  E.g., walks often leave the domain and must be clamped to the boundary; shrinking \(h\) reduces error, but significantly slows down computation (\cref{fig:discrete-bias}). Also, nonlinear functions \(\varphi\) do not in general commute with expectation ($\mathbb{E}[\varphi(X)] \neq \varphi\left(\mathbb{E}[X]\right)$), hence it is not clear how to estimate the function $\exp(-\int_0^{\tau}\zo(X_{t})\diff t)$ in \cref{eq:feynman-kac} in an unbiased way.  Bias is exacerbated in problems with variable diffusion and transport coefficients, which implicitly modify the ideal step size.\discreteSdeFigure{}In contrast, the \(\varepsilon\)-shell in WoS incurs only a miniscule bias at the very end of the walk, leading to far less error overall (\ref{fig:discrete-bias}, \emph{top}); the size of \(\varepsilon\) also has very little effect on performance (\ref{fig:discrete-bias}, \emph{bottom}).  Note that as with ray marching in rendering \citep{Kettunen:2021:Unbiased}, discretized walks do still exhibit fairly predictable and low variance, as long as some bias is tolerable.

\paragraph{Continuous Random Walks} A variety of so-called \emph{grid-free Monte Carlo methods} have been developed for simulating random walks without spatial discretization.  The chief example is the WoS method described in \cref{sec:walk-on-spheres}; variants of this idea include \emph{walk on rectangles}~\cite{Deaconu:2006:RWR} and \emph{walk on boundaries}~\cite{Sabelfeld:2013:RWB}. \citet{Zagajac:1995:FME} describes an alternative grid-free scheme based on shooting rays, albeit only for the basic Laplace equation \(\Delta u = 0\).  WoS has been applied in a limited capacity to problems with piecewise constant coefficients \citep{Maire:2016:PCD,Lejay:2013:PCD}. Ours is the first grid-free method for PDEs with continuously-varying diffusion, transport, and screening coefficients.

\subsection{Volume rendering}

While Monte Carlo solvers have been fairly niche in the PDE community, they are the method of choice for solving the radiative transfer equation in both neutron transport \citep{Spanier:1969:Monte} and more recently rendering \citep{Novak:2018:Monte,Pharr:2016:Physically}.
Of particular interest to us are the so-called \emph{null-scattering} methods, the most simple being delta (or Woodcock)-tracking \cite{Woodcock:1965:Techniques,Coleman:1968:Mathematical,Raab:2008:Unbiased}, where a heterogeneous medium is filled with fictitious ``null-scattering particles'', which transform it into a mathematically equivalent problem where the total density is constant. Early null-scattering methods were physically motivated, but hard to analyze in a mathematical framework. Recent work \citep{Georgiev:2019:Integral, Kutz:2017:Spectral,Galtier:2013:Integral,Miller:2019:Nullscattering} has derived null-scattering versions of the RTE that allow for the construction of new estimators within the same framework, but without the need for a clear physical interpretation. These estimators can be used to exploit known medium properties, for example by introducing control variates \cite{Kutz:2017:Spectral, Novak:2014:Residual}.
More recently, this framework has been built upon further \cite{Jonsson:2020:Direct, Kettunen:2021:Unbiased} and looks to offer a solid foundation for future volumetric rendering work. By expressing our estimators in \cref{sec:algorithms} in the same way, we believe that future work in volumetric rendering will be easily transferable to our Monte Carlo PDE algorithms as well.

\section{Limitations and Future Work}
\label{sec:LimitationsAndFutureWork}

Our methods is not without limitations. Much like volume rendering, variance can increase with increasing values of the parameter $\uCtrl$. Unlike rendering however, the large magnitude of $\zoh({x})$ in \cref{eq:girsanov-pde} (and hence $\uCtrl$ if it is bounding) generally results from derivatives of the transport and diffusion coefficients. In applications involving sharp changes in these coefficients, the derivatives essentially serve as edge detection filters. However, there is plenty of inspiration to take from volume rendering to address this limitation, in particular by bounding the spatially-varying coefficients locally, and by porting other lower variance estimators for the VRE \citep{Novak:2018:Monte,Georgiev:2019:Integral} to solve diffusive PDEs. Furthermore, Monte Carlo techniques not commonly used in rendering but employed in SDE literature such as \emph{Multi-Level Monte Carlo} \citep{Giles:2015:Multilevel} should help accelerate run-time performance. Adaptive weight windows likely provide significant variance reduction \citep{Booth:1984:Importance, Wagner:1998:Automated, Vorba:2016:Adjointdriven}.

Finally, many PDEs, whether constant or variable coefficient, involve Neumann boundary conditions, which remain to be integrated into the walk on spheres framework.

Moving forward, we believe insights from this paper provide building blocks for WoS estimators for an even larger set of PDEs, e.g., with anisotropic coefficients \citep{Boscaini:2016:Anisotropic} or certain nonlinear PDEs. Ultimately, we are optimistic our estimators will open the door to new applications where engineers and scientists can worry less about discretization, and directly analyze models of nearly any complexity---just as in rendering today.

\bibliographystyle{ACM-Reference-Format}
\bibliography{strings-abbrv,SpatiallyVaryingWoS}

\appendix

\section{PDE Transformation: Removing the 1st Order Transport Term}\label{app:remove-transport}

In \cref{sec:our-method}, we reformulate a $2$nd order PDE with a variable transport coefficient into an \emph{equivalent} PDE without the $1$st order term; see \cref{eq:pde-product-rule} and \cref{eq:girsanov-pde}. In this section, we show how to apply a Girsanov transformation to the Feynman-Kac formula introduced in \cref{sec:feynman-kac-formula} to achieve this PDE transformation.

\subsection{The Girsanov Transformation}\label{app:girsanov}

In stochastic calculus, the Girsanov transformation describes how to change the dynamics of a random process (e.g., a diffusion process) under a change of probability measure. Consider a process $X_t$ driven by the SDE $\diff X_t = \fo(X_t) \diff t + \diff W_t$. Then, the Feynman-Kac formula
\begin{multline}\label{eq:feynman-kac-transport}
    u(x) = \mathbb{E}\left[\int_0^{\tau} \euler^{-\int_0^{t}\zo(X_{s}) \diff s}\ \src(X_t) \diff t\ +\right. \\ \left. \euler^{-\int_0^{\tau}\zo(X_{t}) \diff t}\bc(X_{\tau}) \,\middle\vert\, X_0 = x\right]
\end{multline}
expresses the solution to the following PDE as a conditional expectation over random trajectories of $X_t$ inside $\Omega$:
\begin{align}
    \frac{1}{2} \Delta u + \fo({x}) \cdot \nabla u - \zo({x}) u &= -\src({x}) \text{ on } \Omega,\\
    u &= \bc({x}) \text{ on } \partial \Omega \nonumber
\end{align}
The Girsanov transformation makes it possible to change the transport coefficient $\fo(x)$ of the random process $X_t$ by introducing an \emph{importance sampling weight} $Z$ into the Feynman-Kac formula in \cref{eq:feynman-kac-transport} to keep its solution unchanged. In particular, to simulate $X_t$ as a Brownian random walk $W_t$, this transformation provides an \emph{equivalent} Feynman-Kac formulation:
\begin{multline}\label{eq:feynman-kac-girsanov}
    u(x) = \mathbb{E}\left[\int_0^{\tau} \euler^{-\int_0^{t}\zo(W_{s}) \diff s}\ Z(W_t)\ \src(W_t) \diff t\ +\right. \\ 
    \left. \euler^{-\int_0^{\tau}\zo(W_{t}) \diff t}\ Z(W_{\tau})\ \bc(W_{\tau}) \,\middle\vert\, W_0 = x \right]
\end{multline}
where the importance weight $Z$ is explicitly given by the expression:
\begin{equation}
    Z(W_t) := \euler^{\int_0^t \fo(W_s) \cdot \diff W_s\ -\ \frac{1}{2}\int_0^t |\fo(W_s)|^2 \diff s}.
\end{equation}
The term $\int_0^t \fo(W_s) \cdot \diff W_s$ is called a \emph{stochastic integral} since it is defined with respect to variations of a Brownian random process $W_s$. We refer the reader to \citet{Oksendal:2003:Stochastic} for a formal definition of this integral and its properties. We describe how to evaluate this integral next.

\subsection{The Chain Rule of Stochastic Calculus}\label{app:ito}

The stochastic calculus counterpart of the chain rule, known as It$\hat{o}$'s lemma, serves as a useful tool for evaluating stochastic integrals. Given a twice differential function $\gamma(x) : \mathbb{R}^n \mapsto \mathbb{R}$, It$\hat{o}$'s lemma states that its differential $\diff \gamma$, as a function of a Brownian process $W_s$, is given by:
\begin{equation}
    \diff \gamma(W_s) = \nabla \gamma(W_s) \cdot \diff W_s\ +\ \frac{1}{2} \Delta \gamma(W_s) \diff s.
\end{equation}
Integrating this expression over time and rearranging the terms, we get:
\begin{equation}
    \textstyle\int_0^t \nabla \gamma(W_s) \cdot \diff W_s = \gamma(W_t) - \gamma(W_0) - \textstyle\int_0^t \frac{1}{2} \Delta \gamma(W_s) \diff s.
\end{equation}
This integrated version of It$\hat{o}$'s lemma allows us to re-express the importance weight $Z$ without any stochastic integral. In particular, if the transport coefficient $\fo(x)$ takes the form of a gradient of a scalar field $\nabla \gamma(x)$, then an alternative expression for $Z$ is given by:
\begin{equation}
    Z(W_t) = \euler^{\gamma(W_t)\ -\ \gamma(W_0)\ -\ \frac{1}{2} \int_0^t \left(\Delta \gamma(W_s)\ +\
    |\nabla \gamma(W_s)|^2\right) \diff s}.
\end{equation}
\subsection{Derivation of \cref{eq:girsanov-pde}}\label{app:stochastic-pde-derivation}

With this new expression for $Z$, the modified Feynman-Kac formula in \cref{eq:feynman-kac-girsanov} takes the form:
\begin{multline}
    u(x) = \euler^{-\gamma(x)}\ \mathbb{E}\left[\int_0^{\tau} \euler^{-\int_0^{t}\zoh(W_{s}) \diff s}\ \srch(W_t) \diff t\ + \right. \\ \left. 
    \euler^{-\int_0^{\tau}\zoh(W_t) \diff t}\bch(W_{\tau}) \,\middle\vert\, W_0 = x
    \right],
\end{multline}
where 
\begin{align*}
    \srch(x) := \euler^{\gamma(x)} \src(x), \qquad \bch(x) := \euler^{\gamma(x)} \bc(x),\\
    \zoh(x) := \zo(x) + \frac{1}{2}\left(\Delta \gamma(x) + |\nabla \gamma(x)|^2\right).
\end{align*}
Setting $U := \euler^{\gamma(x)} u$, the PDE corresponding to this Feynman-Kac formula is given by:
\begin{align}
    \frac{1}{2}\Delta U - \zoh(x) U &= -\srch(x) \text{ on } \Omega,\\
    U &= \bch(x) \text{ on } \partial \Omega \nonumber,
\end{align}
which does not contain a $1$st order transport term. In \cref{app:girsanov}, we assume $\so(x) := 1$ since it is always possible to absorb a scalar valued diffusion coefficient into the source function $\src(x)$ and screening coefficient $\zo(x)$, as shown with the PDE in \cref{eq:pde-product-rule}. Pattern matching further with \cref{eq:pde-product-rule}, notice that we recover the PDE in \cref{eq:girsanov-pde} by setting the scalar field $\gamma(x)$ to $\frac{1}{2}\ln(\so(x))$.

Finally, we point out that the Girsanov transformation does not only apply when the transport coefficient $\fo(x)$ equals just the gradient of a single scalar field. For instance, to use this transformation, $\fo(x)$ in \cref{eq:pde} can be defined additively as the gradient of several scalar fields, \emph{i.e.}, $\sum_i \nabla \gamma_i(x)$.

\end{document}


\title{Grid-Free Monte Carlo for PDEs with Spatially Varying Coefficients - Supplemental}

\maketitle

\newcommand{\rLT}{r_{-}} 
\newcommand{\rGT}{r_{+}} 

\section{Green's Functions and Poisson Kernels}\label{sec:greens-functions}

Here we provide the Green's function $G^{\zo}(x, y)$ and Poisson kernel $P^{\zo}(x, z)$ for a constant coefficient screened Poisson equation on a ball $\ball(c)$ in $2$D and $3$D. These quantities are needed to estimate the integral expression in Eq. $26$ we derive for variable coefficient PDEs in the paper. Expressions for $\nabla_{x} G^{\uCtrl}(x, y)$ and $\nabla_{x} P^{\uCtrl}(x, z)$ are provided as well to estimate the spatial derivative of Eq. $26$. We also describe how to draw samples $y$ inside $\ball(c)$ from a probability density $p^{\ball}$ that is proportional to the Green's function. Derivations of $G^{\zo}_{2D}$ and $G^{\zo}_{3D}$ can be found in \citet{Duffy:2015:Green}.

\subsection{Centered Expressions}\label{sec:centered}

Assume that the point $x$ lies at the center of a ball $\ball(c)$ with radius $R$, and let $r := |y - x|$. Then the Green's function on $\ball(x)$ in two and three dimensions is given by:
\begin{align}
    G^{\zo}_{2D}(x, y) &= \frac{1}{2\pi} \underbrace{\left(K_0(r\sqrt{\zo}) - \frac{K_0(R\sqrt{\zo})}{I_0(R\sqrt{\zo})} I_0(r\sqrt{\zo})\right)}_{Q^{\zo}_{2D}(r)},\\
    G^{\zo}_{3D}(x, y) &= \frac{1}{4\pi} \sqrt{\frac{2\sqrt{\zo}}{\pi r}} \left(K_{\frac{1}{2}}(r\sqrt{\zo}) - \frac{K_{\frac{1}{2}}(R\sqrt{\zo})}{I_{\frac{1}{2}}(R\sqrt{\zo})} I_{\frac{1}{2}}(r\sqrt{\zo})\right) \nonumber\\
    &= \frac{1}{4\pi} \underbrace{\left(\frac{e^{-r\sqrt{\zo}}}{r} - \frac{e^{-R\sqrt{\zo}}}{R}\left(\frac{\sinh(r\sqrt{\zo})}{r\sqrt{\zo}}\frac{R\sqrt{\zo}}{\sinh(R\sqrt{\zo})}\right)\right)}_{Q^{\zo}_{3D}(r)}, \nonumber
\end{align}
where $I_n$, $I_{n+\frac{1}{2}}$ and $K_n$, $K_{n+\frac{1}{2}}$ (for $n = 0, 1, 2, ...$) denote \emph{modified Bessel functions} of the first and second kind (resp.). Routines to efficiently evaluate these functions are available in numerical libraries such as \emph{Boost} \cite{Schaling:2014:Boost} and \emph{SciPy} \cite{Virtanen:2019:Scipy}.

To compute the probability density $p^{\ball}(x, y) := G^{\zo}(x, y)/|G^{\zo}(x)|$ associated with these Green's functions, we need to evaluate the integrated value of $G^{\zo}$ over all $y$ on $\ball(x)$:
\begin{align}
    |G^{\zo}_{2D}(x)| &:= \int_{B(x)} G^{\zo}_{2D}(x, y) \diff{y}\ =\ \frac{1}{\zo} \left(1 - \frac{1}{I_0(R \sqrt{\zo})}\right),\\
    |G^{\zo}_{3D}(x)| &:= \int_{B(x)} G^{\zo}_{3D}(x, y) \diff{y}\ =\ \frac{1}{\zo} \left(1 - \frac{R\sqrt{\zo}}{\sinh(R \sqrt{\zo})}\right). \nonumber
\end{align}
The Poisson kernel is defined as the normal derivative of the Green's function along the boundary, i.e., for any point $z$ on $\partial \ball(x)$, $P^{\zo}(x, z) := \nabla_{z} G^{\zo}(x, z) \cdot \vec{n}(z)$. In two and three dimensions it is given by:
\begin{align}
    P^{\zo}_{2D}(x, z) &= \frac{1}{2\pi R} \left(\frac{1}{I_0(R \sqrt{\zo})}\right),\\
    P^{\zo}_{3D}(x, z) &= \frac{1}{4\pi R^2} \left(\frac{R\sqrt{\zo}}{\sinh(R\sqrt{\zo})}\right). \nonumber
\end{align}
Notice that in both dimensions, the Poisson kernel equals $\frac{1 - \zo |G^{\zo}(x)|}{|\partial \ball(x)|}$. We exploit this property of the Poisson kernel to develop the delta tracking variant of WoS described in Sec. $5.1$ of the paper.

\subsection{Off-centered Expressions}\label{sec:offcentered}

The next-flight variant of WoS from Sec. $5.2$ in the paper requires off-centered versions of the Green's function and Poisson kernel. In particular, assume $x$ is an arbitrary point inside $\ball(c)$, and let $\rLT := \min(|x - c|, |y - c|)$ and $\rGT := \max(|x - c|, |y - c|)$. Furthermore, let $\theta$ define the angle between the vectors $x - c$ and $y - c$ in $2$D or $3$D. Then the off-centered Green's function on $\ball(c)$ is given by the infinite series:
\begin{align}
    &\begin{multlined}
        G^{\zo}_{2D}(x, y) =
        \frac{1}{2\pi} \sum_{n=-\infty}^{\infty} \cos(n \theta)\ I_n(\rLT\sqrt{\zo})\\ \left(K_n(\rGT\sqrt{\zo}) - \frac{K_n(R\sqrt{\zo})}{I_n(R\sqrt{\zo})} I_n(\rGT\sqrt{\zo})\right),
    \end{multlined}\\
    &\begin{multlined}
        G^{\zo}_{3D}(x, y) = \frac{1}{4\pi} \sum_{n=0}^{\infty} (2n + 1)\ P_n(\cos(\theta)) \left(\sqrt{\frac{\pi}{2 \rLT\sqrt{\zo}}} I_{n+\frac{1}{2}}(\rLT \sqrt{\zo})\right)\\ \sqrt{\frac{2 \sqrt{\zo}}{\pi \rGT}} \left(K_{n+\frac{1}{2}}(\rGT\sqrt{\zo}) - \frac{K_{n+\frac{1}{2}}(R\sqrt{\zo})}{I_{n+\frac{1}{2}}(R\sqrt{\zo})} I_{n+\frac{1}{2}}(\rGT\sqrt{\zo})\right),
    \end{multlined} \nonumber
\end{align}
where $P_n$ denotes the recursively defined Legendre polynomials. As usual, the Poisson kernel can be computed by evaluating $\nabla_{z} G^{\zo}(x, z) \cdot \vec{n}(z)$ on $\partial \ball(c)$. We recover the expressions for $G^{\zo}$ and $P^{\zo}$ in \cref{sec:centered} when $x$ coincides with the ball center $c$.

\begin{figure}[t]
    \centering
    \includegraphics{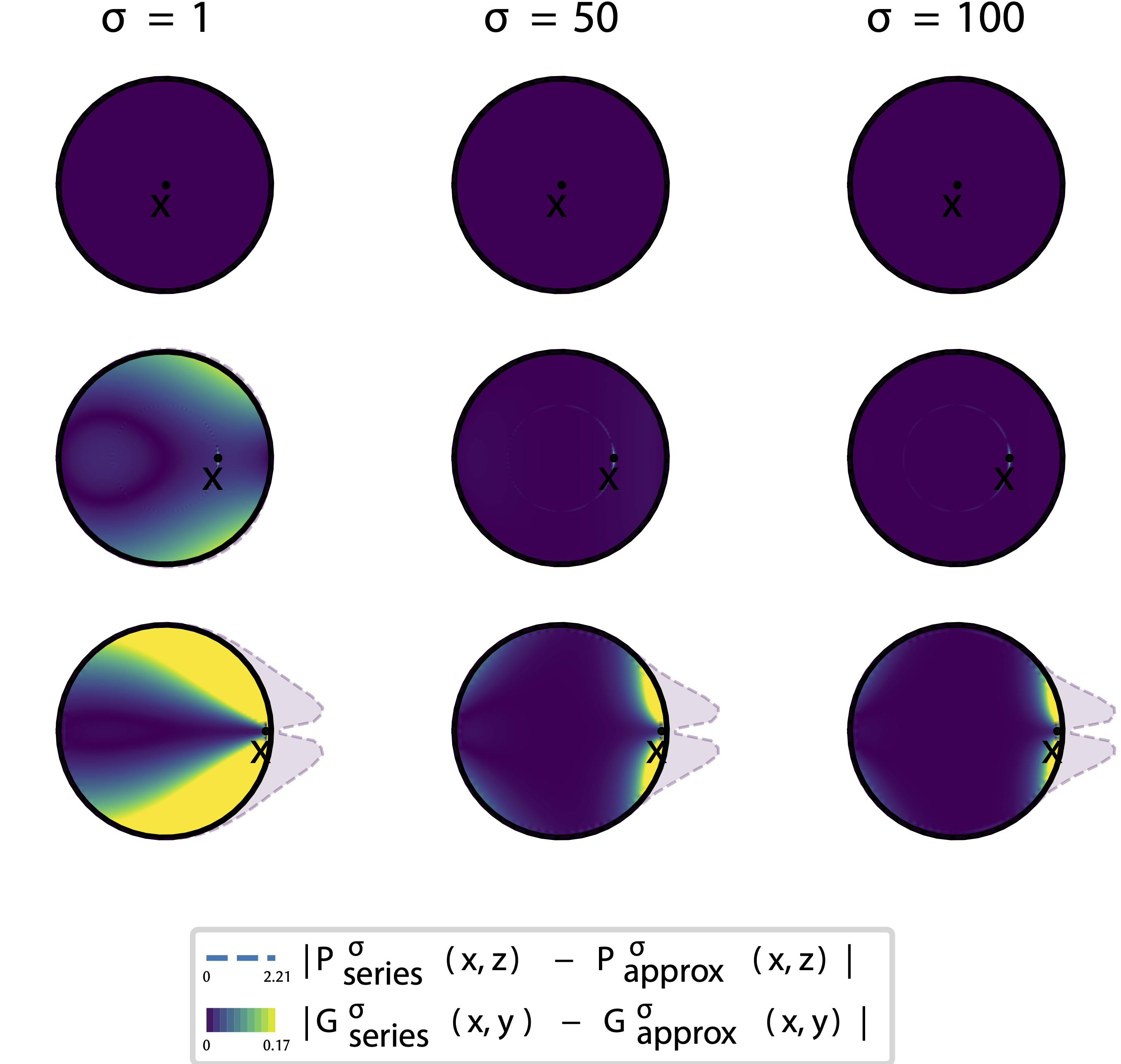}
    \caption{\emph{First Row}: The series and approximate expressions for the Green's function and Poisson kernel on a ball $\ball(c)$ from \cref{sec:offcentered} match exactly when $x$ lies at the center of the ball. \emph{Remaining Rows}: The approximate expressions begin to diverge slightly as $x$ is moved closer to $\partial \ball(c)$, and the value of $\zo$ is decreased.}
    \label{fig:approx-green}
\end{figure}

In practice, we observe that $100$ to $200$ terms are required to accurately approximate these series. To avoid this computational burden, we provide approximations for these off-centered quantities. In particular, let $\vec{u} := x - c$, $\vec{v} := y - c$ and $\vec{w} := y - x$. Then in two and three dimensions we have:
\begin{align}
    G^{\zo}_{2D}(x, y) &= \frac{1}{2\pi}\left(
    Q^{\zo}_{2D}(|\vec{w}|) -
    Q^{\zo}_{2D}\left(\frac{R^2 - \vec{u}\cdot\vec{v}}{R}\right)\right),\\
    G^{\zo}_{3D}(x, y) &= \frac{1}{4\pi}\left(
    Q^{\zo}_{3D}(|\vec{w}|) -
    Q^{\zo}_{3D}\left(\frac{R^2 - \vec{u}\cdot\vec{v}}{R}\right)\right), \nonumber\\
    P^{\zo}_{2D}(x, y) &= \frac{1}{2\pi}\left(
    V^{\zo}_{2D}(|\vec{w}|)\frac{|\vec{v}|^2 - \vec{u}\cdot\vec{v}}{|\vec{w}||\vec{v}|} +
    V^{\zo}_{2D}\left(\frac{R^2 - \vec{u}\cdot\vec{v}}{R}\right)\frac{\vec{u}\cdot\vec{v}}{R|\vec{v}|}\right), \nonumber\\
    P^{\zo}_{3D}(x, y) &= \frac{1}{4\pi}\left(
    V^{\zo}_{3D}(|\vec{w}|)\frac{|\vec{v}|^2 - \vec{u}\cdot\vec{v}}{|\vec{w}||\vec{v}|} +
    V^{\zo}_{3D}\left(\frac{R^2 - \vec{u}\cdot\vec{v}}{R}\right)\frac{\vec{u}\cdot\vec{v}}{R|\vec{v}|}\right), \nonumber
\end{align}
where
\begin{align}
    V^{\zo}_{2D}(r) &:= \sqrt{\zo}\left(K_1(r\sqrt{\zo}) + \frac{K_0(R\sqrt{\zo})}{I_0(R\sqrt{\zo})} I_1(r\sqrt{\zo})\right),\\
    V^{\zo}_{3D}(r) &:= \sqrt{\zo}\sqrt{\frac{2\sqrt{\zo}}{\pi r}} \left(K_{\frac{3}{2}}(r\sqrt{\zo}) + \frac{K_{\frac{1}{2}}(R\sqrt{\zo})}{I_{\frac{1}{2}}(R\sqrt{\zo})} I_{\frac{3}{2}}(r\sqrt{\zo})\right)  \nonumber\\
    &\begin{multlined}
    = \frac{\sqrt{\zo}}{r}\left(e^{-r\sqrt{\zo}}\left(1 + \frac{1}{r\sqrt{\zo}}\right) + \right. \\ \left. \frac{e^{-R\sqrt{\zo}}}{\sinh(R\sqrt{\zo})} \left(\cosh(r\sqrt{\zo}) - \frac{\sinh(r\sqrt{\zo})}{r\sqrt{\zo}}\right)\right). \nonumber
    \end{multlined}
\end{align}
These expressions for $G^{\zo}$ and $P^{\zo}$ are exact when $x$ lies at the center of $\ball(c)$, but begin to diverge slightly from the true values as $x$ is moved closer to $\partial \ball(c)$ and the value of coefficient $\zo$ is decreased; see \cref{fig:approx-green}. In our experiments, we observe that these approximate expressions provide sufficiently accurate results with the next-flight variant of WoS with far less compute, especially when the value of $\zo$ if large.

\subsection{Gradient Expressions}\label{sec:centered-derivatives}

In \cref{sec:derivatives} of this document, we provide an integral expression for the gradient of a PDE solution at a point $x$. To estimate the gradient at the center of $\ball(x)$, we need to evaluate the gradients of the Green's function and Poisson kernel. In two and three dimensions they are given by:
\begin{align}
    &\begin{multlined}
        \nabla_{x} G^{\zo}_{2D}(x, y) = \frac{(y - x)\sqrt{\zo}}{2\pi r} \left(K_1(r\sqrt{\zo}) - \frac{K_1(R\sqrt{\zo})}{I_1(R\sqrt{\zo})} I_1(r\sqrt{\zo})\right),
    \end{multlined} \nonumber\\
    &\begin{multlined}
        \nabla_{x}             G^{\zo}_{3D}(x, y) = \frac{(y - x)\sqrt{\zo}}{4\pi r^2} \left(e^{-r\sqrt{\zo}}\left(1 + \frac{1}{r\sqrt{\zo}}\right) -\right. \\ \left. \left(\cosh(r\sqrt{\zo}) - \frac{\sinh(r\sqrt{\zo})}{r\sqrt{\zo}}\right) \left(\frac{e^{-R\sqrt{\zo}}\left(1 + \frac{1}{R\sqrt{\zo}}\right)}{\cosh(R\sqrt{\zo}) - \frac{\sinh(R\sqrt{\zo})}{R\sqrt{\zo}}}\right)\right),
    \end{multlined} \nonumber\\
    &\begin{multlined}
        \nabla_{x} P^{\zo}_{2D}(x, z) = \frac{(z - x)\zo}{2\pi R} \left(\frac{1}{R\sqrt{\zo}\ I_1(R \sqrt{\zo})}\right),
    \end{multlined} \nonumber\\
    &\begin{multlined}
        \nabla_{x} P^{\zo}_{3D}(x, z) = \frac{(z - x) \zo}{4\pi R^2} \left(\frac{1}{\cosh(R\sqrt{\zo}) - \frac{\sinh(R\sqrt{\zo})}{R\sqrt{\zo}}}\right).
    \end{multlined} \nonumber
\end{align}

\subsection{Sampling}\label{sec:sampling-greens}

To sample from the probability density $p^{\ball} := G^{\zo}(x, y)/|G^{\zo}(x)|$ associated with the \emph{centered} Green's functions $G^{\zo}$ in \cref{sec:centered}, we first pick a direction $\vec{y}$ uniformly on the unit sphere \cite{Arvo:2001:SST}. A radius $r$ is then sampled from the distribution $2 \pi r p^{\ball}$ in $2$D, or $4 \pi r^2 p^{\ball}$ in $3$D, using rejection sampling. The extra factor in front of $p^{\ball}$ accounts for the change of measure between polar and Cartesian coordinates. For rejection sampling, we bound the radial density by the following case dependent function:
\begin{align}
    h(R, \zo) &:= \begin{cases}
    \max(2.2*\max(1/R, 1/\zo), 0.6*\max(\sqrt{R}, \sqrt{\zo})) & R \leq \zo,\\
    \max(2.2*\min(1/R, 1/\zo), 0.6*\min(\sqrt{R}, \sqrt{\zo})) & \text{otherwise}.
    \end{cases}
\end{align}
The final sample point is given by $y = r\vec{y} + x$.

Generating samples from an \emph{off-centered} Green's function $G^{\zo}(x, y)$ in \cref{sec:offcentered} is more challenging since we do not know of a closed-form expression for $|G^{\zo}(x)|$. While using a uniform density $\frac{1}{|\ball(c)|}$ for $p^{\ball}$ suffices for unbiased sampling, more sophisticated techniques to generate samples according to the profile of $G^{\zo}$ exist. We recommend the weighted reservoir version of resampled importance sampling provided in \cite[Alg. 3]{Bitterli:2020:Spatiotemporal}.

\section{Spatial Gradient}\label{sec:derivatives}

Applications often require computing not just the solution to a PDE, but the spatial gradient of the solution as well. Fortunately, estimating the gradient $\nabla_{x} u(x)$ of Eq. $26$ from the paper at a point $x$ adds virtually no cost on top of estimating the solution $u(x)$ itself. In particular, either of our WoS algorithms can be used to evaluate the following integral expression for $\nabla_{x} u(x)$ at the center of a ball $\ball(x)$:
\begin{multline}\label{eq:derivative-integral}
    \nabla_{x} u(x) = \frac{1}{\sqrt{\so(x)}}\left( \int_{\ball(x)} \srch(y, \sqrt{\so}\ u) \ \nabla_{x} G^{\uCtrl}(x, y) \diff{y}\ \right. +\\
    \left. \int_{\surf(x)} \sqrt{\so(z)}\ u(z)\ \nabla_{x} P^{\uCtrl}(x, z) \diff{z}\right) \ -\ \frac{u(x)}{2 \so(x)} \nabla_{x} \so(x).
\end{multline}
The value of $\nabla_{x} u(x)$ only needs to be estimated in the first ball in any walk\textemdash the solution estimates it depends on can be computed recursively using the delta tracking or next-flight estimators; see Sec. $5$ in the paper. Furthermore, the parameters $\uCtrl$, $\pdfBall$, $\pdfSurf$, $\probBall$ and $\probSurf$ remain unchanged with either algorithm.

\section{Pseudo-code}\label{sec:pseudocode}

Here we provide pseudo-code for the two variants of walk on spheres presented in Sec. $5$ of the paper. To maintain consistency with the paper , we assume the transport coefficient $\fo({x}) = \vec{0}$ over the entire domain.

\begin{algorithm}[h!]
\KwIn{A point $x \in \Omega$.}
\KwOut{A single sample estimate $\est{u}(x)$ of the solution to Eq. $1$.}

\tcc{Compute distance and closest point to $x$ on $\partial \Omega$}
$d, \overline{{x}} \leftarrow $ DistanceToBoundary($x$)\;

\tcc{Return boundary value $\bc$ at $\overline{x}$ if $x \in \partial \Omega_{\epsilon}$}
\lIf{ $d < \epsilon$ } {
    \Return $\bc(\overline{x})$
}

\tcc{Estimate source contribution at random point $y \in \ball(x)$}
$y \sim \frac{G^{\uCtrl}(x, y)}{\abs{G^{\uCtrl}(x)}}$\;
$\est{\pathsrc} \leftarrow \frac{\abs{G^{\uCtrl}(x)}}{\sqrt{\so(x)\so({y})}}\src({y}) $

\tcc{Decide whether to sample volume or boundary term in Eq. $27$}
\uIf{ $\mu \sim \mathcal{U} \leq \uCtrl \abs{G^{\uCtrl}(x)}$ } {
    \tcc{Estimate solution at $y \in \ball(x)$; adjust estimate by null-event contribution from Eq. $21$}
    \Return $\sqrt{\frac{\so(y)}{\so(x)}}\left(1 - \frac{\zoh(y)}{\uCtrl}\right)\text{ DeltaTrackingEstimate}(y) + \est{\pathsrc}$\;

}
\Else{
    \tcc{Estimate solution at random point $z \in \partial \ball(x)$}
    $z \sim \frac{1}{\abs{\surf(x)}}$\;
    \Return $\sqrt{\frac{\so(z)}{\so(x)}} \text{ DeltaTrackingEstimate}(z) + \est{\pathsrc}$\;
}

\caption{DeltaTrackingEstimate($x$)}
\label{alg:delta-tracking}
\end{algorithm}

\begin{algorithm}[h!]
\KwIn{A point $x \in \Omega$.}
\KwOut{A single sample estimate $\est{u}(x)$ of the solution to Eq. $1$.}

\tcc{Compute distance and closest point to $x$ on $\partial \Omega$}
$d, \overline{{x}} \leftarrow $ DistanceToBoundary($x$)\;

\tcc{Return boundary value $\bc$ at $\overline{x}$ if $x \in \partial \Omega_{\epsilon}$}
\lIf{ $d < \epsilon$ } {
    \Return $\bc(\overline{x})$
}

\tcc{Initialize series expressions from Eq. $29$}
$\est{\pathtr} \leftarrow 0$\;
$\est{\pathsrc} \leftarrow 0$\;

\tcc{Initialize path throughput}
$\pathth \leftarrow 1$\;

\tcc{Sample random exit point $z \in \partial \ball(x)$ used across all entries in $\est{\pathtr}$}
$z \sim \frac{1}{\abs{\surf(x)}}$\;

\tcc{Initialize temporary variable to track current sample point inside $\ball(x)$}
$x_c \leftarrow x$

\While{True}{
    \tcc{Accumulate boundary contribution}
    $\est{\pathtr}$ += $\frac{P^{\uCtrl}(x_c, z)}{p^{\partial \ball}(z)} \pathth$\;

    \tcc{Use path throughput as Russian Roulette probability to terminate loop}
    ${\mathbb{P}}^{\mathrm{RR}} = \min(1, \pathth)$\;
    \lIf{${\mathbb{P}}^{\mathrm{RR}} < \mu \sim \mathcal{U}$ }{
        \textbf{break}
    }
    $\pathth$ /= ${\mathbb{P}}^{\mathrm{RR}}$\;

    \tcc{Sample next random point $x_n \in \ball(x)$}
    $x_n \sim \frac{1}{\abs{\ball(x)}}$\;

    \tcc{Update path throughput}
    $\pathth$ *= $\frac{G^{\uCtrl}(x_c, x_n) (\uCtrl - \zoh(x_n))}{\pdfBall(x_n)}$\;

    \tcc{Accumulate source contribution}
    $\pathsrc$ += $\frac{\src(x_n)}{\sqrt{\so(x_n)} (\uCtrl - \zoh(x_n))} \pathth$\;

    \tcc{Update current sample point inside $\ball(x)$}
    $x_c \leftarrow x_n$
}

\tcc{Estimate solution at $z \in \partial \ball(x)$}
\Return  $\frac{1}{\sqrt{\so(x)}} (\sqrt{\so(z)}\ \est{\pathtr} \text{ NextFlightEstimate}(z) + \est{\pathsrc})$\;

\caption{NextFlightEstimate($x$)}
\label{alg:next-flight}
\end{algorithm}

\bibliographystyle{ACM-Reference-Format}
\bibliography{strings-full,SpatiallyVaryingWoS}